\documentclass[twocolumn,superscriptaddress,aps,amsmath,amssymb]{revtex4}
\usepackage{amsxtra,amssymb,bm,amsbsy,amsmath,amsthm,latexsym,dsfont,array, layout,graphicx,physics,mathrsfs,color,soul,enumerate,footmisc}
\usepackage[colorlinks=true,citecolor=blue,linkcolor=blue,anchorcolor=blue,urlcolor=blue]{hyperref}
\hypersetup{linktoc=all}

\renewcommand{\ketbra}[2]{\!\left|{#1}\rangle\!\langle{#2}\right|}
\renewcommand{\Tr}{{\operatorname{Tr}}}
\newcommand{\E}{\mathrm{E}}
\newcommand{\e}{\epsilon}
\renewcommand{\a}{\bm{a}}
\newcommand{\ba}{\bm{\alpha}}
\newcommand{\A}{\bm{A}}
\newcommand{\tA}{{\widetilde{\A}}}
\renewcommand{\b}{\bm{b}}
\newcommand{\bet}{\bm{\eta}}
\newcommand{\ra}{\rightarrow}

\newcommand{\tot}{\mathrm{tot}}
\newcommand{\T}{{\mathrm{T}}}
\newcommand{\bz}{\bm{\zeta}}
\newcommand{\z}{\zeta}
\renewcommand{\t}{\tau}
\newcommand{\J}{\bm{J}}
\newcommand{\Z}{\bm{Z}}
\newcommand{\D}{\bm{D}}
\newcommand{\ri}{\rm{i}}
\newcommand{\perm}{\rm{perm}}
\renewcommand{\u}{\bm{u}}
\newcommand{\U}{\bm{U}}
\newcommand{\tilu}{\widetilde{\u}}
\renewcommand{\v}{\bm{v}}
\renewcommand{\rm}{\mathrm}
\renewcommand{\S}{{\mathrm{S}}}
\renewcommand{\d}{{\mathrm d}}
\renewcommand{\l}{\!\left}
\renewcommand{\r}{\right}

\newcommand{\blk}{\color{black}}

\begin{document}

\title{Exact dynamics and thermalization of open quantum systems coupled to reservoir through particle exchanges}
\author{Fei-Lei Xiong}
\affiliation{Department of Physics and Center for Quantum Information Science,
National Cheng Kung University, Tainan 70101, Taiwan}
\author{Wei-Min Zhang}
\email{wzhang@mail.ncku.edu.tw}
\affiliation{Department of Physics and Center for Quantum Information Science,
National Cheng Kung University, Tainan 70101, Taiwan}

\begin{abstract}
In this paper, we study the exact dynamics of general open systems interacting with its environment through particle exchanges. The paper includes two main results. First, by taking advantage of the propagating function in the coherent state representation, we solve the exact master equation, whose solution is expressed in terms of the Keldysh nonequilibrium Green functions. Second, in the dynamical perspective, we provide a rigorous  thermalization process of open quantum systems.  
\end{abstract}

\maketitle

\section{Introduction}

In the realistic world, physical systems are inevitably coupled to environments, which makes the theory of open quantum systems vastly used in many fields of physics, chemistry, biology, and engineering. The systematic study of open quantum systems has aroused the researchers' interest since the 1960s~\cite{FV63,S61,Z60,N58,BP02,W12,GP04}, and becomes more and more important for the prosperously developing field of quantum information processing~\cite{GP04,WM09,NC02}, quantum transport theory~\cite{HJ08,YZ16}, and rapidly improving time-resolved measurement technologies~\cite{WBK05,KBS16}. One of the most crucial problems in dealing with open quantum systems is how to determine explicitly the evolution of open quantum system states, through which all the information about the system dynamics can be obtained. However, due to the contamination of the huge environment, the system dynamics is non-unitary and always involves complicated fluctuation and dissipation. As a consequence, up to date, most open quantum systems can only be dealt with perturbative methods, such as Born-Markov approximation or cutting-off in Nakajima-Zwanzig operator projective technique~\cite{N58,Z60,BP02}. Only a few classes of open quantum systems, e.g., the harmonic oscillator in quantum Brownian motion~\cite{CL83,HPZ92} and the open systems interacting with particle-exchange coupling ~\cite{ZLX12}, can an exact master equation be obtained, let alone the exact state evolution.

In a series of papers~\cite{TZ08,JTZY10,LZ12,ZLX12}, we have obtained the general exact master equation for the open quantum systems whose interaction only involves particle exchanges. This large class of open systems is an extension of the famous Fano-Anderson model~\cite{A61,F61} and characterizes many physical phenomena in different systems, such as Fano resonance in atomic and condensed matter physics~\cite{MFK10}, Anderson localization in many-body systems~\cite{A58}, photon-atom bound states in photonic crystal~\cite{Y87,J87,KKS94}, quantum transport in quantum dot junctions~\cite{HJ08}, impurity defects in solids~\cite{M13}, etc.  Although it has been studied for decades, the exact general master equation has not been derived until very recently~\cite{TZ08,JTZY10,LZ12,ZLX12}. With the progress, various perspectives of the open systems can be studied, e.g., memory effects~\cite{YAZ13}, entanglement dynamics~\cite{TZL11,LYZ16,AZ07}, decoherence~\cite{TZ08,LTZ16,YW14}, the fluctuation-dissipation theorem~\cite{ZLX12}, exact transient quantum transport~\cite{YZ18,LXZ15,YZ16}, etc.

In this paper, one main result we obtained is the solution of the exact master equation, i.e., the exact form of the reduced density matrix at an arbitrary later time. Note that the time-dependence of the reduced density matrix includes complete information about the system dynamics, it offers more perspectives in studying the memory effects~\cite{LHW18,BLP16,RHP14}, entanglement dynamics and dynamical phase transition~\cite{H18} of such class of open systems. Using the result of the reduced density matrix, we also study the general thermalization in this paper, which is another important and hot topic for both experimentalists~\cite{TCF12,GKL12} and theorists~\cite{D91,K13,AHZ18,XLZ15,S94,LPS09,SF12,R08,R09,PSSV11,CR10,HCSH18} in recent years. In equilibrium statistic mechanics, the thermal distribution is based on the assumption of equal probability of all permissible microstates~\cite{C98}. However, the foundation of thermalization has always been a tough problem and a long-term goal of physicists~\cite{PSSV11}. In the past decade, by taking advantage of the eigenstate
thermalization hypothesis, researchers study the thermalization of closed quantum systems and make interesting contributions in understanding the physics beneath it~\cite{D91,K13,AHZ18,S94,LPS09,SF12,R08,R09,PSSV11,CR10}. In contrast, in this paper, we discuss the thermalization of open systems from the dynamical perspective. With the exact evolution of the open quantum systems, such an aim is achieved.  

The rest of the paper is organized as follows. In Sec.~\ref{sec_2}, we introduce the system we concerned about and briefly review the previous results related to our present work. In Sec.~\ref{sec_3}, we shall derive the exact solution of the reduced density matrix.  In Sec.~\ref{sec_4},  we explain the physical consequences of the solution and discuss the general thermalization of the open systems. A brief summary will be given in Sec.~\ref{sec_5}.

\section{Overview of the exact master equation}
\label{sec_2}

The system we are interested are characterized by the Hamiltonian $H_{\S}=\sum_{i,j=1}^d \e_{ij} a^\dag_{i} a_j$, and interacts with bosonic (fermionic) environment described by $H_{\E}=\sum_{k} \e_{k} b^\dag_{k} b_k$, via the interaction 
$H_{\rm{int}}=\sum_{j,k}(V_{jk} a^\dag_{j} b_k+\rm{H.c.})$, where $a_j^\dag$ ($a_j$) is the creation (annihilation) operator of the $j$th level in the system; $b_k^\dag$ ($b_k$) is the creation (annihilation) operator of the $k$-mode in the environment; $\e_{ij}$ and $\e_k$ characterize the energy levels of the system and the environment, and  $V_{jk}$ is the interaction strength between them. If the creation and annihilation operators satisfy the commutation (anti-commutation) relations, the corresponding systems is bosonic (fermionic).

To make the notations more compact, we denote that $\a^\dag=(\!\begin{array}{ccc}
a^\dag_1  & \cdots & a^\dag_d\end{array}\!)$, $\a=(\!\begin{array}{ccc}
a_1 & \cdots & a_d \end{array}\!)^{\T}$, and the matrix constituted of ${\e}_{ij}$ as $\bm{\e}_\S$. 
The spectrum of the environment and the system-environment interaction can be specified with the spectral density function $\bm{J}(\e)$, whose elements are defined through
$J_{ij}(\omega)=2\pi \sum_{k} V_{k i}
V^*_{k j}\delta(\e-\e_k)$. 
We assume that initially the system and the environment is decoupled~\cite{LCD87}, i.e.,  
$\rho_{{\rm{tot}}}(0)=\rho(0)\otimes\rho_{\E}(0)$, where $\rho(0)$ is an arbitrary physical state of the system and $\rho_{\E}(0)=\otimes_{k}e^{ -\beta (\e_k-\mu)b_k^\dag b_k} /{\cal Z}$ is the thermal state of the environment with inverse temperature $\beta$ and chemical potential $\mu$. Here, 
${\cal Z}=\prod_k |1\mp e^{-\beta[\e_k-\mu]}|^{\mp 1}$ denotes 
the grand partition function of the environment with the upper and lower sign corresponding to the bosonic and fermionic case respectively.  Hereafter, we also use this convention.

The reduced density matrix at latter time $t$ is connected to the initial state 
 through the propagating function ${\cal{J}} (\bet^{\ast},\bet,t;\bz'^{\ast},\bz,0)$ in the coherent state representation, i.e.,
\begin{align}
\langle \bet^{\ast}\vert \rho(t)\vert \bet\rangle =\int\d\mu(\bz)\d\mu(\bz')\langle {\bz}^{\ast}\vert \rho(0)\vert {\bz}^{\prime}\rangle
{\cal{J}} (\bet^{\ast},\bet,t;\bz'^{\ast},\bz,0)\, ,\label{eq_ifl}
\end{align}
where $\bz^{ \ast}=(\!\begin{array}{ccc}
\z^{ \ast}_1  & \cdots & \z^{ \ast}_d\end{array}\!)$ and $ \bz=(\!\begin{array}{ccc}
\z^{ }_1 & \cdots & \z^{ }_d \end{array}\!)^{\T}$,   with the components being complex (Grassmannian) for bosons (fermions);
$\bra{\bz^*}=\bra{0}e^{\bz^* \bm{a}}$ ($\ket{\bz}=e^{\bm{a}^\dag \bz}\ket{0}$)
is the unnormalized coherent state with $\bra{0}$ ($\ket{0}$) standing for the vacuum state~\cite{ZFG90}; $\d\mu(\bz)$ is the integral measure, and
$\d\mu(\bz)=\prod_{j=1}^d ({\d^2\z_j}/{\pi}) e^{-\z^{*}_j \z_j}$ for bosons and $\d\mu(\bz)=\prod_{j=1}^d {\d \z^{*}_j \d\z_j} e^{-\z^{*}_j \z_j}$ for fermions. 
With coherent state path-integral approach, the propagating function can be derived as~\cite{JTZY10,LZ12}
\begin{align}
{\cal{J}} (\bet^{\ast},\bet,t;\bz'^{\ast},\bz,0) =\!\frac{e^{{\bet}^{\ast}\frac{\v}{1\pm \v}{\bet}}}{ \l\vert  1\pm \v \r\vert^{\pm 1}} e^{\pm {\bz}^{\prime\ast}\tA\bz+{\bz}^{\prime\ast}\tilu^\dag{\bet}+{\bet}^{\ast}\tilu \bz}\,,
\label{eq_inf}
\end{align}
where $\tA=1-\u^{\dag}\frac1{1 \pm \v}\u$, $\tilu =  \frac1{1\pm \v} \u$, with the matrices $\u$ and $\v$ being respectively the abbreviations of the nonequilibrium Green functions $\u(t)$ and $\v(t,t)$. More specifically, $\u(t)$ is the spectral Green function with elements defined as $u_{ij}(t)=\langle [a_i(t),a_j^\dag(0)]_\mp\rangle$~\cite{K18}, and
 $\v(t,t)$ is related to the system's particle number originated from the environment, reading~\cite{ZLX12} 
\begin{align}
\v(t,t) = \int^t_{0}\! \d t_1 \! \int^t_0\!
\d t_2 \, \u(t-t_1)\, \widetilde{\boldsymbol{g}}(t_1, t_2) \,
\u^\dag(t-t_2) \,.  \label{eq_v1}
\end{align} 
Here, 
$\widetilde{\boldsymbol{g}}(t_1,t_2) =
\int\frac{\d\e}{2\pi}f(\e) \boldsymbol{J}(\e)
e^{-i\e(t_1-t_2)}$ denotes a system-bath correction,
with $f(\e)=\frac1{e^{\beta(\e-\mu)} \mp 1}$ standing for the initial particle distribution of the environment.

Taking advantage of Eqs.~\eqref{eq_ifl}-\eqref{eq_inf}, the master equation has been derived~\cite{TZ08,JTZY10,LZ12,ZLX12}, reading
\begin{align}\label{eq_r1}
\frac{d\rho(t)}{dt} = & \frac1{\ri}\Big[\widetilde{H}_\S(t),
\rho(t)\Big]  + \sum_{ij} \Big\{\gamma_{ij}(t)
\Big[2a_j\rho(t)a^{\dag}_i \nonumber \\
& - a^{\dag}_ia_j\rho(t) -\rho(t) a^{\dag}_ia_j\Big] +
\widetilde{\gamma}_{ij}(t)\Big[a^\dag_i \rho(t)
a_j \nonumber \\ & \pm a_j\rho(t)a^{\dag}_i \mp a^{\dag}_ia_j\rho(t) 
- \rho(t) a_ja^{\dag}_i \Big] \Big\}\,.
\end{align}
In this master equation, $\widetilde{H}_\S(t)=\sum_{ij}
\widetilde{\e}_{ij}(t)a^\dag_i a_j$ is the renormalized system Hamiltonian; $\gamma_{ij}(t)$ and $\widetilde{\gamma}_{ij}(t)$ characterize the dissipation and fluctuation rate induced by the system's coupling with the environment, respectively. The coefficients $\widetilde{\e}_{ij}(t)$, $\gamma_{ij}(t)$ and $\widetilde{\gamma}_{ij}(t)$ are determined only by the non-equilibrium Green functions $\u(t)$ and $\v(t,t)$~\cite{ZLX12}.

Furthermore, the Green function $\u(t)$ satisfies the Dyson equation  and the solution can be expressed as~\cite{ZLX12}
\begin{align}\label{eq_u}
\u(t)=\sum_l \Z_l  e^{-\ri \e_l t} + \int%
\frac{\d \e}{2 \pi} \D(\e) e^{- \ri \e t}\,.
\end{align}
The quantities $\e_l$, $\Z_l$ and $\D(\e)$ are all connected with the Green function in the energy domain 
\begin{align}
\U(z) = \frac{1}{z{\bm{\rm{I}}} - \bm{\e}_\S-\bm{\Sigma}(z)}\,,
\end{align}
where $\bm{\Sigma}(z) =\int \frac{\d\e}{2\pi} 
\frac{\boldsymbol{J}(\e)}{z-\e}$ is the self-energy correction. 
$\e_l$ is the energy of the $l$th localized mode and is determined by the singularities of $\U(z)$, i.e., it satisfies the equation $\det(\e_l-\bm{\e}_\S -\bm{\Sigma}(\e_l))=0$. 
$\Z_l$ is the Hermitian matrix characterizing the amplitude of the $l$th localized mode, reading $\Z_l =\frac{1}{2\pi \ri} \oint_{C_l}\!\d z\,\U(z)$\,, 
where $C_l$ is the positively-oriented curve in the neighbourhood of $z=\e_l$. 
$\D(\e)$ is the spectrum of the system broadened by the environment 
and reads $\bm{D}(\e)=\U(\e+\ri 0^+) \J(\e) \U(\e-\ri 0^+)$~\footnote{The $\Z_l $'s and $\D(\e)$ are the discrete and continuous components of the modified spectrum of the system $\D_\S(\e)$, respectively. They are connected through the relation $\D_\S(\e)=2\pi\sum_l \Z_l \delta(\e-\e_l)+\D(\e)$, where $\D_\S(\e)$ is defined through $\D_\S(\e):=-2\Im[\U(\e+\ri 0^+)]$ (See Ref.~\cite{ZLX12}).}.

\section{The exact system state evolution}
\label{sec_3}

With the initial system state $\rho(0)$, and the Green functions $\u(t)$ and $\v(t,t)$, the reduced density matrix $\rho(t)$ at arbitrary instant can be determined from Eq.~\eqref{eq_r1} in principle. However, because the coefficients are all time-dependent, the problem is too complicated to solve directly. Luckily, one can solve it by taking advantage of the propagating function in Eqs.~\eqref{eq_ifl}-\eqref{eq_inf}, as is shown in this section. 
First, in Sec.~\ref{sec_3a}, we shall introduce some conventions and derive $\rho(0)$'s elements in the coherent state representation. In Sec.~\ref{sec_3b}, we obtain $\rho(t)$'s elements at arbitrary later time. 
In Sec.~\ref{sec_3c}, the exact form of $\rho(t)$ will be given. 

\subsection{Representation of the initial system state}
\label{sec_3a}
To obtain the density matrix elements at arbitrary time, one needs to carry out the integrals of Eq.~\eqref{eq_ifl} with the explicit expression of $\langle {\bz}^{\ast}\vert \rho(0)\vert {\bz}^{\prime}\rangle$.
Define the Fock state 
\begin{align}
\ket{I} =\frac{\a_I^{\dag}}{\sqrt{i_1 !\cdots i_d !}}\ket{0}, \quad  \bra{I}=\bra{0} \frac{\a_{I^\T}}{\sqrt{i_1 !\cdots i_d !}}
\,.
\end{align}
where $\a_I^{\dag}:=(a_1^{\dag})^{i_1}\cdots (a_d^{\dag})^{i_d}$, $\a_{I^\T}:=(a_d)^{i_d}\cdots  (a_1)^{i_1}$, and $i_n$ denotes the particle number in the $n$th level. For bosons, the $i_n$ takes the values $0,1,2,\cdots$, while for fermionic systems, $i_n$ is either $0$ or $1$. Corresponding to the state $\ket{I}$, we define
a class of sequences with the form 
\begin{align}
I=(\underbrace{1,\cdots, 1}_{i_1},\underbrace{2,\cdots, 2}_{i_2},\cdots, \underbrace{d, \cdots, d}_{i_d})\,. 
\end{align}
Then the initial system density matrix can be generally expressed as
\begin{align}\label{eq_rho0}
\rho(0)
&=\sum_{IJ}\rho_{IJ}(0)\frac{\a_I^{\dag} \ketbra{0}{0} \a_{J^\T} }{\sqrt{i_1 !\cdots i_d ! j_1 ! \cdots j_d !}}\,,
\end{align}
where the summation is over all the possible physical pairs of $I$ and $J$. 
That is, except for the constraints over $I$ and $J$ individually, for fermions and massive bosons, the sequences $I$ and $J$ together satisfy the constraint $i_1+\cdots+i_d=j_1+\cdots+j_d$. With Eq.~\eqref{eq_rho0} and the definition of coherent states, it is easy to find 
\begin{align}\label{eq_init}
\langle {\bz}^{\ast}\vert \rho(0)\vert {\bz}^{\prime}\rangle
&=\sum_{IJ}\rho_{IJ}(0)\frac{\bz^*_I \bz'_{J^{\T}}}{\sqrt{i_1 !\cdots i_d ! j_1 ! \cdots j_d !}}\,,
\end{align}
where $\bz^*_I=(\z_1^{\ast})^{i_1}\cdots (\z_d^{\ast})^{i_d}$ and $\bz'_{J^{\T}}=(\z'_d)^{j_d}\cdots (\z'_1)^{j_1}$.

\subsection{Evolution of the density matrix elements in the coherent state representation}
\label{sec_3b}


Following Eqs.~\eqref{eq_inf} and ~\eqref{eq_init},
Eq.~\eqref{eq_ifl} can be re-expressed as
\begin{align}
\langle \bet^{\ast}\vert \rho&(t)\vert \bet\rangle =\frac{e^{{\bet}^{\ast}\frac{\v}{1\pm \v}{\bet}}}{ \l\vert  1\pm \v \r\vert^{\pm 1}} \sum_{IJ}\frac{\rho_{IJ}(0)}{\sqrt{i_1 !\cdots i_d ! j_1 ! \cdots j_d !}}\nonumber\\
&\int\!\d\mu(\bz)\d\mu(\bz') {\bz^*_I \bz'_{J^{\T}}}  e^{\pm {\bz}^{\prime\ast}\tA\bz+{\bz}^{\prime\ast}\tilu^\dag{\bet}+{\bet}^{\ast}\tilu \bz}\,.\label{eq_denmatelm}
\end{align}
In order to obtain the explicit form of $\langle \bet^{\ast}\vert \rho(t)\vert \bet\rangle$, one needs to simplify the integral.  The result is given in the following (See Appendix~\ref{app_a} for the details). 

\subsubsection{Bosonic case}
In the bosonic case, the result is
\begin{align}\label{eq_exp_bos}
\sum_{I'J'} \perm(\tA_{J',I'})(\bet^{\ast}\tilu)_{\overline{I'}}(\tilu^\dag \bet)_{\overline{J'}^{\T}}\,, 
\end{align}
where we have used the following conventions:
\begin{enumerate}[(i)]    
    \item $I'=(\underbrace{1,\cdots, 1}_{i'_1},\cdots, \underbrace{d, \cdots, d}_{i'_d})$ is a subsequence of $I$ and $\overline{I'}=(\underbrace{1,\cdots, 1}_{\overline{i'_1}},\cdots, \underbrace{d, \cdots, d}_{\overline{i'_d}})$ is its complement, with $i_n=i'_n+\overline{i'_n}$ for each $n$.  By subsequence, we mean deleting some elements in a sequence and keeping the remaining elements in the original order. 
    The summation is over all the possible ways of choosing $I'$ and $J'$ in $I$ and $J$, respectively, with each pair satisfying the constraint $i'_1+\cdots +i'_d=j'_1+\cdots +j'_d$; 
    \item  $\tA_{J',I'}$ is the matrix obtained by first taking $j'_n$ copies of the $n$th column of $\tA$ in Eq.~\eqref{eq_inf} for each $n$ and then taking $i'_n$ copies of the $n$th row of the matrix obtained by the first step~\cite{TDH13};
\item $\perm(\tA_{J',I'})$ stands for the \emph{permanent} of the matrix $\tA_{J',I'}$~\cite{TDH13};
    \item 
    $(\bet^* \tilu)_{\overline{I'}}:=[(\bet^* \tilu)_1]^{\overline{i'_1}}\cdots [(\bet^* \tilu)_d]^{\overline{i'_d}}$, $(\tilu^\dag \bet)_{\overline{J'}^{ \rm T}}:=[(\tilu^\dag \bet)_d]^{\overline{j'_d}}\cdots [(\tilu^\dag \bet)_1]^{\overline{j'_1}}$. 
\end{enumerate}

\subsubsection{Fermionic case}
In the fermionic scenario, the integral can be simplified to
\begin{align}\label{eq_exp_fer}
\sum_{I'J'} \delta^I_{I'\overline{I'}} \delta^J_{J'\overline{J'}}\det(\tA_{J',I'})(\bet^{\ast}\tilu )_{\overline{I'}}(\tilu^\dag \bet)_{\overline{J'}^{\T}}\,.
\end{align}
Here, we have used the conventions:
\begin{enumerate}[(i)]    
    \item $I'$, $\overline{I'}$, $J'$, $\overline{J'}$ and the summation follow the same conventions as those in the bosonic case, except that all the elements in the sequences appear at most once; 
    \item  $I'\overline{I'}$ is the new sequence obtained by joining $I'$ and $\overline{I'}$ end-to-end. 
    $\delta^I_{I'\overline{I'}}=\pm 1$ when $I' \overline{I'}$ is an even/odd permutation of $I$; 
    \item 
    $\tA_{J',I'}$ follows the same convention as in the bosonic case. However, as all the elements $1,\cdots,d$ appear at most once in $I'$ and $J'$, $\tA_{J',I'}$ is in fact a submatrix of $\tA$~\cite{TDH13};
    \item $\det(\tA_{J',I'})$ stands for the \emph{determinant} of the matrix $\tA_{J',I'}$;  
    \item $(\bet^* \tilu)_{\overline{I'}}$ and $(\tilu^\dag \bet)_{\overline{J'}^{ \rm T}}$ follow the same convention as in the bosonic case.  
\end{enumerate}

\subsubsection{Common expression}
The results in the previous two cases can be expressed as a common expression. By denoting 
\begin{equation}
s_\pm=\left\{
\begin{array}{lr}
1\qquad  & (\text{boson})\,,\\ 
\delta^I_{I'\overline{I'}} \delta^J_{J'\overline{J'}}\qquad  & (\text{fermion})\,,
\end{array}
\right.
\end{equation}
and 
\begin{equation}
|\tA_{J',I'}|_\pm:=
\left\{
\begin{array}{lr}
\perm(\tA_{J',I'})\qquad  & (\text{boson})\,,\\ 
\det \big(\tA_{J',I'}\big)\qquad  & (\text{fermion})\,,   
\end{array}
\right.
\end{equation}
Eqs.~\eqref{eq_exp_bos} and~\eqref{eq_exp_fer} can be summarized as
\begin{align}\label{eq_int}
\sum_{I'J'}s_\pm  |\tA_{J',I'}|_\pm (\bet^{\ast}\tilu )_{\overline{I'}}(\tilu^\dag \bet)_{\overline{J'}^{\T}}\,.
\end{align}
Thus, in the coherent state representation, the elements of $\rho(t)$ can be given by
\begin{align}
\langle \bet^{\ast}\vert \rho(t)\vert \bet\rangle = &\sum_{IJ}\frac{\rho_{IJ}(0)}{\sqrt{i_1 !\cdots i_d ! j_1 ! \cdots j_d !}}\nonumber \\
\sum_{I'J'}s_\pm & |\tA_{J',I'}|_\pm (\bet^{\ast}\tilu )_{\overline{I'}} \frac{e^{{\bet}^{\ast}\frac{\v}{1\pm \v}{\bet}}}{ \l\vert  1\pm \v \r\vert^{\pm 1}}  (\tilu^\dag \bet)_{\overline{J'}^{\T}}\,.\label{eq_rhot_coh}
\end{align}

\subsection{Exact form of the density matrix}
\label{sec_3c}

The reduced density matrix $\rho(t)$ satisfying Eq.~\eqref{eq_rhot_coh} can be explicitly written as
\begin{align}
\rho(t)=\sum_{IJ}{\rho_{IJ}(0)}\sum_{I'J'}s_\pm \frac{|\tA_{J',I'}|_\pm(\a^{\dag}\tilu )_{\overline{I'}} \rho^{{\rm{th}}}(t) (\tilu^\dag \a)_{ \overline{J'}^{\T}}}{\sqrt{i_1 !\cdots i_d ! j_1 ! \cdots j_d !}} \,,\label{eq_rhot}
\end{align}
where $\rho^{{\rm{th}}}(t)=\!\frac{e^{ \a^{\dag}\ln(\frac{\v}{1\pm \v}) \a}}{\vert 1\pm \v\vert^{\pm 1}}$ is a thermal-like state~\cite{XLZ15}; $(\a^{\dag} \tilu)_{\overline{I'}}:=[(\a^{\dag} \tilu)_1]^{\overline{i'_1}}\cdots [(\a^{\dag} \tilu)_d]^{\overline{i'_d}}$ and $(\tilu^\dag \a)_{\overline{J'}^{ \rm T}}:=[(\tilu^\dag \a)_d]^{\overline{j'_d}}\cdots [(\tilu^\dag \a)_1]^{\overline{j'_1}}$. 
Equation~\eqref{eq_rhot} is directly obtained from  Eq.~\eqref{eq_rhot_coh} with the identity
\begin{align}
\langle \bet^{\ast}\vert (\a^{\dag}\tilu )_{\overline{I'}}e^{\a^{\dag}\ln(\frac{\v}{1\pm \v}) \a}&(\tilu^\dag \a)_{ \overline{J'}^{\T}}\vert \bet\rangle\nonumber\\
=&(\bet^{\ast}\tilu )_{\overline{I'}}e^{\bet^{\ast}\frac{\v}{1\pm \v}\bet}(\tilu^\dag \bet)_{ \overline{J'}^{\T}}\,,\label{eq_proof}
\end{align}
which can be easily derived with the properties of coherent states that
\begin{subequations}
\begin{align}
\langle \bet^{\ast}\vert (\a^{\dag}\tilu )_{\overline{I'}}=\langle \bet^{\ast}\vert (\bet^{\ast}\tilu )_{\overline{I'}}\,,\\ 
(\tilu^\dag \a)_{ \overline{J'}^{\T}}\vert \bet\rangle =(\tilu^\dag \bet)_{ \overline{J'}^{\T}}\vert \bet\rangle\,,
\end{align}
\end{subequations}
and  
\begin{align}
\langle \bet^{\ast}\vert e^{\a^{\dag}\ln(\frac{\v}{1\pm \v}) \a}\vert \bet\rangle =e^{\bet^{\ast}\frac{\v}{1\pm \v}\bet}\,. 
\end{align}

\section{Physical interpretation of the solution and the thermalization}
\label{sec_4}

In this section, we shall discuss the physics beneath the solution. The physical interpretation of the solution will be given in Sec.~\ref{sec_4a}, and the thermalization problem will be discussed in Sec.~\ref{sec_4b}.

\subsection{Physical interpretation of the solution}
\label{sec_4a}

To explain the physical consequence contained in Eq.~\eqref{eq_rhot}, we consider two limiting cases first. One is that there is no particle in the environment initially, 
and the other is no particle in the system initially. 
Finally, we shall consider the joint effect and the general solution.

\subsubsection{No particle in the environment initially}
\label{sec_4a1}
In this case, the environment is initially in a vacuum state, so that $f(\e)=0$ and  
$\v(t,t)=0$. Consequently, in Eq.~\eqref{eq_rhot}, $\tA=1-\u^{\dag}\frac1{1 \pm \v}\u$, $\tilu =  \frac1{1\pm \v} \u$, and $\rho^{{\rm{th}}}(t)=\frac1{\vert 1\pm \v\vert^{\pm 1}}e^{ \a^{\dag}\ln(\frac{\v}{1\pm \v}) \a}$ reduces to $\A=1-\u^{\dag}\u$, $\u$, and $\rho^{{\rm{th}}}(t)=\vert 0\rangle \langle 0\vert$, respectively. 
As a result, $\rho(t)$ is simply reduced to  
\begin{align}
\label{eq_rhodis}
\rho(t)=\sum_{IJ}{\rho_{IJ}(0)}\sum_{I'J'} s_{\pm}
\frac{|A_{J',I'}|_\pm(\a^{\dag}\u)_{\overline{I'}} \ketbra{0}{0}\! (\u^\dag  \a)_{ \overline{J'}^{\T}}}{\sqrt{i_1 !\cdots i_d ! j_1 ! \cdots j_d !}} \,.
\end{align}
Compared with the form of the initial state of  Eq.~\eqref{eq_rho0}, one can see that the factor $\a_I^{\dag} \ketbra{0}{0} \a_{J^\T}$ evolves to $\sum_{I'J'} s_{\pm}|A_{J',I'}|_\pm (\a^{\dag}\u)_{\overline{I'}}\ketbra{0}{0}\!(\u^\dag  \a)_{ \overline{J'}^{\T}}$.  Note that $\u$ quantifies the particles maintained in the system and $\A=1-\u^\dag \u$ quantifies their loss into the environment. Therefore, the summation describes all the possibilities that part of the system particles maintains in the system while the others dissipate into the environment. That is, the solution in Eq.~\eqref{eq_rhodis} precisely describes the pure \emph{dissipation} process.  
The sign $s_{\pm}$ and the quantity $|A_{J',I'}|_\pm$ are the manifestations of the particle exchange symmetry,  which is quite similar to the cases encountered in the boson and fermion sampling~\cite{AA11}.  

\blk


\subsubsection{No particle in the system initially}
\label{sec_4a2}
In this case, the system initial state reads $\rho(0)=\ketbra{0}{0} $, i.e., all the $\rho_{IJ}(0)$'s are $0$ except the one that $I$ and $J$ are both empty sequence. Following Eq.~\eqref{eq_rhot}, one can easily find that
\begin{align}\label{eq_rho_f}
\rho(t)=\rho^{{\rm{th}}}(t)=\frac{e^{ \a^{\dag}\ln(\frac{\v}{1\pm \v}) \a}}{\vert 1\pm \v\vert^{\pm 1}}\,.
\end{align}
For understanding the physical consequence of Eq.~\eqref{eq_rho_f}, one needs to consider the physical meaning of $\v(t,t)$. We introduce the spectral Green function of the total system, i.e., $\bm{u}_\tot(t)=e^{-\ri {\bm{\e}_\tot} t}$, where $\bm{\e}_\tot$ is the energy matrix of the total system Hamiltonian. Formally, $\bm{u}_\tot(t)$ can be written in matrix blocks, i.e.,
\begin{align}
{\u}_\tot(t)=\l(\begin{array}{cc}
{\u}_{{\rm{SS}}}(t) & {\u}_{{\rm{SE}}}(t)\\
{\u}_{{\rm{ES}}}(t) & {\u}_{{\rm{EE}}}(t)
\end{array}\r)\,,  
\end{align}
where $\u_{\rm {XY}}(t)$ is the spectral Green function between X and Y, with 
S (E) being the abbreviation of the system (environment). 
Equivalent to Eq.~\eqref{eq_v1}, $\v(t,t)$ can be expressed as
\begin{align}\label{eq_v}
\v(t,t)={\u}_{{\rm{SE}}}(t) f(\bm{\e}_\E) {\u}_{{\rm{ES}}}^{\dag}(t)\,,
\end{align} 
where $f(\bm{\e}_\E)$ is the particle distribution of the environment. 
From the equation, it is obvious that $\v(t,t)$ characterizes the average particle number transported from all the energy levels of the environment to the system. Therefore, $\rho^{{\rm{th}}}(t)$ is a thermal-like state completely contributed by the environmental particles propagating to the system, with the average particle number characterized by $\v(t,t)$. In other words, the solution~\eqref{eq_rho_f} comes from the thermal fluctuation process. 

\subsubsection{Joint effect between fluctuation and dissipation}
Now we consider the general result of Eq.~\eqref{eq_rhot}. From Sec.~\ref{sec_4a1}, we can conclude that particles initially in the system contribute to the terms $s_\pm$, $|\A_{J',I'}|_\pm$, $(\a^{\dag}\u)_{\overline{I'}}$ and $(\u^\dag  \a)_{ \overline{J'}^{\T}}$. From Sec.~\ref{sec_4a2}, we know that $\rho^{\rm{th}}(t)$ originates from the particles initially in the environment. When initially particles coexist in both the system and environment, the dissipation property is modified due to the fluctuation. That is, $\A=1-\u^\dag \u$ and $\u$ 
are modified as $\tA=1-\u^\dag \frac1{1\pm \v} \u$ and $\tilu=\frac1{1\pm \v} \u$, 
respectively. Due to the effect $\A\rightarrow \tA$, the probability of the system particles dissipating into the environment becomes larger for bosons, while it becomes less for fermions. Correspondingly, the modification  $\u\rightarrow\tilu$ reveals that, for bosons, the amplitude of the system particles maintaining in the system becomes smaller; while for fermions, it becomes larger. 
These are due to the statistic properties of identical particles. 
For bosons, if an environmental level is occupied with some particles, the probability of the system particles hopping into that level becomes larger, as is described in Feynman lectures~\cite{F11} and manifested in many phenomena such as superradience and Bose-Einstein condensation. While for fermions, due to the Pauli exclusion principle, if an environmental level is occupied, the transition onto it is forbidden.

\subsection{The thermalization}
\label{sec_4b}
In this subsection, we shall study the asymptotic behavior of the state as the time approaches infinity. The absence or presence of localized modes determines whether the system can be finally thermalized, so we consider these two cases separately.  

In the case that there are no localized modes (see Appendix~\ref{app_b}), $\u(t)$ and $\v(t,t)$ would finally evolve to
\begin{subequations}
\begin{align}
&\u(\infty)=0\,,\\
&\v(\infty,\infty)=\int \frac{\d \e}{2 \pi} f(\e) \bm{D}(\e):=\overline{\bm{n}}\,.\label{eq_vinf}
\end{align}\label{eq_uv}
\end{subequations}
In this case, the final state of the system would be (See Appendix~\ref{app_c} for the details)
\begin{align}
\label{eq_rho_infty}
\rho(\infty)=\rho^{{\rm{th}}}(\infty)=\frac{e^{ \a^{\dag}\ln(\frac{\overline{\bm{n}}}{1\pm \overline{\bm{n}}}) {\a}}}{\vert 1\pm \overline{\bm{n}}\vert^{\pm 1}} \,.
\end{align}
Equation~\eqref{eq_rho_infty} implies that the final particle distribution in the system is completely characterized by the matrix  
$\overline{\bm{n}}$~\cite{SR15}, i.e.,
\begin{align} 
\Tr[ \rho(\infty) a_j^\dag a_i ]=\overline{\bm{n}}_{ij}\,.
\end{align}
With the expression of $\overline{\bm{n}}$ in Eq.~\eqref{eq_vinf} and the properties of the spectral function $\D(\e)$ (that it is positive-semidefinite and $\int \frac{\d \e}{2 \pi} \bm{D}(\e)=\bm{\rm{I}}$), the final particle distribution can be seen as a weighted sum of the Bose/Fermi distribution. That is, without localized modes, the system would finally reach a thermal-like state, instead of the conventional thermal state.

When the coupling strength between the system and the environment is \emph{very weak}, 
then the spectral density $\J(\e)$ and the Lamb shift $\bm{\Delta}(\e)$ both tend to vanish, i.e.,
\begin{align}\label{eq_JDel}
\J(\e)\ra \bm{0}\,,\qquad  \bm{\Delta}(\e) \ra \bm{0}\,.    
\end{align}
 Following 
 \begin{align}\label{eq_DS}
\!\bm{D}(\e)
\!=\!\frac{1}{\e\!-\!\bm{\e}_\S\!-\!\bm{\Delta}(\e)\!+\!\!\frac{\ri\bm{J}(\e)}{2}} \J(\e) \frac{1}{\e\!-\!\bm{\e}_\S\!-\!\bm{\Delta}(\e)\!-\!\frac{\ri\bm{J}(\e)}{2}},
\end{align}
and the careful analysis in Appendix~\ref{app_d}, one can find that under  condition~\eqref{eq_JDel}, 
\begin{align}\label{eq_D_app}
\bm{D}(\e)\ra 2\pi \delta (\e \bm{\rm{I}}-\bm{\e}_\S)\,.
\end{align}
That is, when the system-environment coupling becomes very weak, the broadening and Lamb shift of the system energy levels also vanishes, making the spectrum of the system converging to that of the isolated system. 
As a consequence of Eqs.~\eqref{eq_D_app} and~\eqref{eq_vinf}, $\overline{\bm{n}}$  approaches to the conventional Bose/Fermi distribution, i.e., 
\begin{align}
\overline{\bm{n}}\ra
f(\bm{\e}_\S)=\frac{1}{e^{\beta(\bm{\e}_\S-\mu)}\mp 1}\,.
\end{align}
Thus, Eq.~\eqref{eq_rho_infty} converges to 
\begin{align}
\rho(\infty)= {e^{-\beta\a^{\dag}(\bm{\e}_\S-\mu) {\a}}}/\Tr[e^{-\beta\a^{\dag}(\bm{\e}_\S-\mu) {\a}}]\,,
\end{align}
which is exactly the thermal state of the system $H_{\S}= \a^\dag \bm{\e}_\S \a$ in the grand canonical ensemble with inverse temperature $\beta$ and chemical potential $\mu$ of the environment. This provides a rigorous proof that in the weak-coupling limit, the exact evolution of an open quantum system would reproduce in the steady state limit the thermal state in conventional statistic mechanics. 

On the other hand, if there are localized modes, their contribution to the oscillations in $\u(t)$ (See Eq.~\eqref{eq_u}) would survive as $t$ approaches infinity, i.e., 
\begin{align}\label{eq_ulimit}
{\u}(t\rightarrow \infty) =  \sum_l {\bm{Z}}_l  & e^{-\ri\e_l t}\,.
\end{align}
 Following the expression of the reduced density matrix in Eq.~\eqref{eq_rhot}, the final state must be expressed in terms of the coefficients $\rho_{IJ}(0)$, i.e., the system keeps the memory of its initial state. Therefore, the system cannot be thermalized. 

\blk

\section{Summary}
\label{sec_5}
In this paper, we have investigated a general solution of open quantum systems interacting with the environment through particle exchanges. The exact evolution of the reduced density matrix is given in terms of the nonequilibrium Green functions. We explained the physical consequences of the solution. With the exact density matrix, we study the thermalization process. 
We obtain the result of equilibrium statistical mechanics from the dynamical perspective and go beyond it. That is, when there are no localized modes and the system-environment coupling is very weak, the final state would be as expected from the conventional statistical mechanics; for no localized modes but relatively strong  coupling regime, the steady state would be thermal-like, which departures from the prediction of conventional statistical mechanics; when there are localized modes, the system keeps the memory of the initial state and can not be thermalized.

With the explicit expression of the reduced density matrix, one can obtain the complete information about the system dynamics, which is quite important for the rapidly developing quantum thermodynamics and quantum information, because their central physical quantity, entropy, is directly related to the state. It is also noteworthy that the model studied in our work involves non-Markovian nature. With the explicit form of the density matrix evolution, one can study the memory effects from more perspectives, e.g., quantum coherence, entanglement, and dynamical phase transition. Although we only consider the single-reservoir case, our result can be directly extended to the multi-reservoir case by just extending the corresponding expressions of the nonequilibrium Green functions $\u(t)$ and $\v(t,t)$ to multi-reservoirs (multi-leads in nano/quantum devices). Therefore, it is also easy to apply to quantum transport theory.

\acknowledgements
We thank Yu-Wei Huang, Matisse Wei-Yuan Tu and Li Li for helpful discussions. This work is supported by the Ministry of Science and Technology of the Republic of China under the Contracts No. MOST 107-2811-M-006-534 and No. MOST 108-2811-M-006-518.

\begin{appendix}
\begin{widetext}

\section{Simplification of the integral}
\label{app_a}

\subsection{Bosonic case}
\label{app_a1}
From Eq.~\eqref{eq_denmatelm}, for bosons,
\begin{align}
&\int\d\mu(\bz)\d\mu(\bz') \bz^*_I\bz'_{J^{\T}} \! e^{ {\bz}^{\prime\ast}\tA\bz+{\bz}^{\prime\ast}\tilu^\dag{\bet}+{\bet}^{\ast}\tilu \bz}\nonumber\\
=&\int\prod_{n=1}^d \frac{\d^2\z_n \d^2\z'_n}{\pi^2} \bz^*_I\bz'_{J^{\T}}{\exp\left\{ -\l(\!\begin{array}{cc}
    \bz^{\ast} & \bz^{\prime\ast}\end{array}\!\r)\l(\!\begin{array}{cc}
    1 & 0\\
    - \tA & 1
    \end{array}\!\r)\l(\!\begin{array}{c}
    \bz\\
    \bz'
    \end{array}\!\r)+\l(\!\begin{array}{cc}
    \bz^{\ast} & \bz^{\prime\ast}\end{array}\!\r)\l(\!\begin{array}{c}
    \bm{0}\\
    \tilu^\dag \bet
    \end{array}\!\r)+\l(\!\begin{array}{cc}
    \bet^{\ast}\tilu  & \bm{0}\end{array}\!\r)\l(\!\begin{array}{c}
    \bz\\
    \bz'
    \end{array}\!\r)\right\} }\nonumber\\
=&\eval{\partial_{\ba_{J^{{\rm T}}}^{\ast}\ba_{I}}\int\prod_{n=1}^d \frac{\d^2\z_n \d^2\z'_n}{\pi^2}
    \exp\left\{ -\l(\!\begin{array}{cc}
    \bz^{\ast} & \bz^{\prime\ast}\end{array}\!\r)\l(\!\begin{array}{cc}
    1 & 0\\
    - \tA & 1
    \end{array}\!\r)\l(\!\begin{array}{c}
    \bz\\
    \bz'
    \end{array}\!\r)+\l(\!\begin{array}{cc}
    \bz^{\ast} & \bz^{\prime\ast}\end{array}\!\r)\l(\!\begin{array}{c}
    \ba\\
    \tilu^\dag \bet
    \end{array}\!\r)+\l(\!\begin{array}{cc}
    \bet^{\ast}\tilu  & \ba^{\ast}\end{array}\!\r)\l(\!\begin{array}{c}
    \bz\\
    \bz'
    \end{array}\!\r)\right\} }_{\ba,\ba^{\ast}=0}\nonumber\\
=&\eval{\partial_{\ba_{J^{{\rm T}}}^{\ast}\ba_{I}}e^{ \ba^{\ast}\tA\ba+\bet^{\ast}\tilu \ba+\ba^{\ast}\tilu^\dag \bet} }_{\ba,\ba^{\ast}=0}\,,
\end{align}
where we have used the convention $\partial_{\ba_{J^{{\rm T}}}^{\ast}\ba_{I}}:=(\partial_{\ba_d^{\ast}})^{j_d}\cdots(\partial_{\ba_1^{\ast}})^{j_1} (\partial_{\ba_1})^{i_1}\cdots(\partial_{\ba_d})^{i_d}$ and the formula of Gaussian integral~\cite{KL09}.

$\partial _{\ba_{J^{\T}}^*\ba_{I}}e^{\ba^*\tA\ba+\ba^{\ast}\tilu^\dag \bet+\bet^{\ast}\tilu \ba}|_{\ba=\ba^{\ast
}=0}$ is only related to the coefficient of $\alpha_d^{i_d}\cdots\alpha_1^{i_1} \alpha_1^{\ast
j_1} \cdots \alpha_d^{*j_d}$ in the polynomial expansion of  $e^{\ba^*\tA\ba+\ba^{\ast}\tilu^\dag \bet+\bet^{\ast}\tilu \ba}$. Note 
\begin{align}
e^{\ba^*\tA\ba+\ba^{\ast}\tilu^\dag \bet+\bet^{\ast}\tilu \ba 
}=\sum_{k_1,k_2,k_3=0}^{\infty }\frac{(\ba^*\tA\ba 
) ^{k_1}}{k_1!}
\frac{( \bet^{\ast}\tilu \ba) ^{k_2}}{k_2!}
\frac{(\ba^{\ast}\tilu^\dag \bet)^{k_3}}{k_3!}\,.\label{eq_expansion}
\end{align}
The terms with factor $\alpha_d^{i_d}\cdots\alpha_1^{i_1} \alpha_1^{\ast
j_1} \cdots \alpha_d^{*j_d}$ can be obtained through
that 
\begin{enumerate}[(i)]    
\item $\frac{( \ba^*\tA\ba) ^{i_1^{\prime
}+\cdots +i_d'}}{( i_1'+\cdots +i_d^{\prime
}) !}$ contributes to $\alpha_d^{i'_d}\cdots\alpha_1^{i'_1} \alpha_1^{\ast
j'_1} \cdots \alpha_d^{*j'_d}$,
\item $\frac{( \bet^{\ast}\tilu \ba) ^{\overline{i_1'}%
+\cdots +\overline{i_d'}}}{( \overline{i_1'}%
+\cdots +\overline{i_d'}) !}$ contributes to $\alpha_d^{\overline{i_d'}} \cdots \alpha_1^{
\overline{i_1'}}$,
\item $\frac{( \ba^{\ast}\tilu^\dag \bet) ^{\overline{j_1'}%
+\cdots +\overline{j_d'}}}{( \overline{j_1'}%
+\cdots +\overline{j_d'}) !}$ contributes to $\alpha_1^{*\overline{j_1'}} \cdots \alpha
_d^{*\overline{j_d'}}$, 
\end{enumerate}
where $i'_1,\cdots,i'_d, j'_1,\cdots,j'_d, \overline{i'_1},\cdots,\overline{i'_d}, \overline{j'_1},\cdots,\overline{j'_d}$ satisfy the constraint 
\begin{subequations}
\label{eq_app_con_bsn}
\begin{align}
&i'_1,\cdots,i'_d, j'_1,\cdots,j'_d, \overline{i'_1},\cdots,\overline{i'_d}, \overline{j'_1},\cdots,\overline{j'_d}\in \{0,1,2,\cdots\}\,; \\
&\overline{i_1'}=i_1-i_1',\quad \cdots,\quad  \overline{i_d'}=i_d-i_d'\,; \\
&\overline{j_1'}=j_1-j_1', \quad \cdots, \quad  \overline{j_d'}=j_d-j_d'\,;\\
&i_1'+\cdots
+i_d'=j_1'+\cdots +j_d'\,.  
\end{align}
\end{subequations}
 Note,
\begin{enumerate}[(i)]
    \item the coefficient of $\alpha_d^{i'_d}\cdots\alpha_1^{i'_1} \alpha_1^{\ast
j'_1} \cdots \alpha_d^{*j'_d}$ in $\frac{( \ba^*\tA\ba 
) ^{i_1'+\cdots +i_d'}}{( i_1^{\prime
}+\cdots +i_d') !}$ is $\frac{\perm( \tA%
_{J',I'}) }{i_1'!\cdots i_d^{\prime
}!j_1'!\cdots j_d'!}$; 
\item the coefficient of $\alpha_d^{\overline{i_d'}}\cdots
\alpha_1^{\overline{i_1'}}$ in $\frac{( \bet^{\ast}\tilu \ba) ^{\overline{i_1'}+\cdots +\overline{i_d^{\prime
}}}}{( \overline{i_1'}+\cdots +\overline{i_d'}%
) !}$ is $\frac{[(\bet^{\ast}\tilu)_1]^{
\overline{i_1'}} \cdots [(\bet^{\ast}\tilu)_d]^{\overline{i_d'}}}   {\overline{{i_1'}}!\cdots \overline{{i_d'}}!}=\frac{(\bet^{\ast}\tilu)_{\overline{I'}}}{\overline{{i_1'}}!\cdots \overline{{i_d'}}!}$;
\item the coefficient of $\alpha_1^{*\overline{j_1'}}\cdots
\alpha_d^{*\overline{j_d'}}$ \bigskip in $\frac{(
\ba^{\ast}\tilu^\dag \bet) ^{\overline{j_1'}+\cdots +\overline{j_d'}}}{( \overline{j_1'}+\cdots +\overline{j_d'}) !}$ is $\frac{[(\tilu^\dag \bet)_d]^{\overline{j_d'}%
}\cdots [(\tilu^\dag \bet)_1]^{\overline{j_1'}}}{\overline{{j_1'}}%
!\cdots \overline{{j_d'}}!}=\frac{(\tilu^\dag \bet)_{\overline{J'}^{\rm{T}}}}{\overline{{j_1'}}%
!\cdots \overline{{j_d'}}!}$. 
\end{enumerate}
Therefore, the coefficient of $\alpha_d^{i_d}\cdots\alpha_1^{i_1} \alpha_1^{\ast
j_1} \cdots \alpha_d^{*j_d}$ 
in $\frac{
( \ba^*\tA\ba) ^{i_1'+\cdots
+i_d'}}{( i_1'+\cdots +i_d') !}
\frac{( \ba^{\ast}\tilu^\dag \bet) ^{\overline{j_1'}+\cdots +
\overline{j_d'}}}{( \overline{j_1'}+\cdots +
\overline{j_d'}) !}\frac{( \bet^{\ast}\tilu \ba) ^{
\overline{i_1'}+\cdots +\overline{i_d'}}}{( 
\overline{i_1'}+\cdots +\overline{i_d'}) !}$ is $\frac{\perm( \tA_{J',I'}) }{i_1^{\prime
}!\cdots i_d'!j_1'!\cdots j_d'!}\frac{(\bet^{\ast}\tilu)_{\overline{I'}}}{\overline{{i_1'}}!\cdots \overline{{i_d'}}!}\frac{(\tilu^\dag \bet)_{\overline{J'}^{\rm{T}}}}{\overline{{j_1'}}%
!\cdots \overline{{j_d'}}!}$,  
and the total coefficient of $\alpha_d^{i_d}\cdots\alpha_1^{i_1} \alpha_1^{\ast
j_1} \cdots \alpha_d^{*j_d}$ 
in Eq.~\eqref{eq_expansion} 
is
the summation of all these terms with $i_{( \cdot ) }'$'s $j_{(
\cdot ) }'$'s satisfying Eq.~\eqref{eq_app_con_bsn}, i.e.,
\begin{align}
\sum_{i_1',\cdots ,i_d',j_1',\cdots
,j_d'}\frac{\perm( \tA_{J',I'}) }{i_1^{\prime
}!\cdots i_d'!j_1'!\cdots j_d'!}\frac{(\bet^{\ast}\tilu)_{\overline{I'}}}{\overline{{i_1'}}!\cdots \overline{{i_d'}}!}\frac{(\tilu^\dag \bet)_{\overline{J'}^{\rm{T}}}}{\overline{{j_1'}}%
!\cdots \overline{{j_d'}}!}\,.
\end{align}
Also note that 
\begin{align}
\partial _{\ba_{J^{\T}}^*\ba_{I}}( \alpha_d^{i_d}\cdots\alpha_1^{i_1} \alpha_1^{\ast
j_1} \cdots \alpha_d^{*j_d}) =i_1!\cdots i_d!j_1!\cdots j_d!\,,
\end{align}
therefore
\begin{align}
\eval{\partial _{\ba_{J^{\T}}^*\ba_{I}}e^{\ba^*\tA%
\ba+\ba^{\ast}\tilu^\dag \bet+\bet^{\ast}\tilu \ba}}_{\ba=\ba^*=0}
=\sum_{i_1',\cdots,i_d',j_1',\cdots,j_d'}C_{i_1'}^{i_1}\cdots C_{i_d^{\prime}}^{i_d}C_{j_1'}^{j_1}\cdots C_{j_d^{\prime}}^{j_d}\perm(\tA_{J',I'})(\bet^{\ast}\tilu)_{\overline{I'}}(\tilu^\dag\bet)_{\overline{J'}^{\rm{T}}}\,,
\end{align}
where 
$C_{k}^{n}=\frac{n!}{k!( n-k) !}$ stands for the binomial coefficient. The factor $C_{i_1'}^{i_1}\cdots C_{i_d^{\prime
}}^{i_d}C_{j_1'}^{j_1}\cdots C_{j_d'}^{j_d}$ is
the number of ways of obtaining $I'$ from $I$ as well as
obtaining $J'$ from $J$. So we can transform the summation over
all the possible $I'$'s and $J'$'s to the summation of all
the possible ways of obtaining subsequences $I'$'s and $J'$'s from $I$ and $J$, respectively. Therefore, the factor $C_{i_1^{\prime
}}^{i_1}\cdots C_{i_d'}^{i_d}C_{j_1^{\prime
}}^{j_1}\cdots C_{j_d'}^{j_d}$ is absorbed in the sum and the
result can be reexpressed as
\begin{align}
\int\d\mu(\bz)\d\mu(\bz') \bz^*_I\bz'_{J^{\T}} \! e^{ {\bz}^{\prime\ast}\tA\bz+{\bz}^{\prime\ast}\tilu^\dag{\bet}+{\bet}^{\ast}\tilu \bz}=\sum_{I',J'}\perm( \tA_{J^{\prime
},I'}) (\bet^{\ast}\tilu)_{\overline{I'}}(\tilu^\dag\bet)_{\overline{J'}^{\rm{T}}}\,.
\end{align}

\subsection{Fermionic case}
\label{app_a2}
From Eq.~\eqref{eq_denmatelm}, for fermions, we have
\begin{align}
&\int\d\mu(\bz)\d\mu(\bz') \bz^*_I\bz'_{J^{\T}} e^{- {\bz}^{\prime\ast}\tA\bz+{\bz}^{\prime\ast}\tilu^\dag{\bet}+{\bet}^{\ast}\tilu \bz}\nonumber\\
=&\int\prod_{n=1}^d {(\d\z_n^*\d\z_n \d\z'^*_n \d\z'_n)} \bz^*_I\bz'_{J^{\T}} \! \exp\left\{ -\l(\!\begin{array}{cc}
    \bz^{\ast} & \bz^{\prime\ast}\end{array}\!\r)\l(\!\begin{array}{cc}
    1 & 0\\
     \tA & 1
    \end{array}\!\r)\l(\!\begin{array}{c}
    \bz\\
    \bz'
    \end{array}\!\r)+\l(\!\begin{array}{cc}
    \bz^{\ast} & \bz^{\prime\ast}\end{array}\!\r)\l(\!\begin{array}{c}
    \bm{0}\\
    \tilu^\dag \bet
    \end{array}\!\r)+\l(\!\begin{array}{cc}
    \bet^{\ast}\tilu  & \bm{0}\end{array}\!\r)\l(\!\begin{array}{c}
    \bz\\
    \bz'
    \end{array}\!\r)\right\}\nonumber\\
=&\eval{\partial_{\ba_{J^{{\rm T}}}^{\ast}\ba_{I}}\int{\prod_{n=1}^d {(\d\z_n^*\d\z_n \d\z'^*_n \d\z'_n)}}
    \exp\left\{ -\l(\!\begin{array}{cc}
    \bz^{\ast} & \bz^{\prime\ast}\end{array}\!\r)\l(\!\begin{array}{cc}
    1 & 0\\
     \tA & 1
    \end{array}\!\r)\l(\!\begin{array}{c}
    \bz\\
    \bz'
    \end{array}\!\r)+\l(\!\begin{array}{cc}
    \bz^{\ast} & \bz^{\prime\ast}\end{array}\!\r)\l(\!\begin{array}{c}
    \ba\\
    \tilu^\dag \bet
    \end{array}\!\r)+\l(\!\begin{array}{cc}
    \bet^{\ast}\tilu  & \ba^{\ast}\end{array}\!\r)\l(\!\begin{array}{c}
    \bz\\
    \bz'
    \end{array}\!\r)\right\} }_{\ba,\ba^{\ast}=0}\nonumber\\
=&\eval{\partial_{\ba_{J^{{\rm T}}}^{\ast}\ba_{I}}e^{- \ba^{\ast}\tA\ba+\bet^{\ast}\tilu \ba+\ba^{\ast}\tilu^\dag \bet} }_{\ba,\ba^{\ast}=0}\,,
\end{align}
where we have used the convention $\partial_{\ba_{J^{{\rm T}}}^{\ast}\ba_{I}}:=(\partial_{\alpha_d^{\ast}})^{j_d}\cdots(\partial_{\alpha_1^{\ast}})^{j_1} (\partial_{\alpha_1})^{i_1}\cdots(\partial_{\alpha_d})^{i_d}$ and the formula of Grassmannian Gaussian integral~\cite{KL09}.

$\partial _{\ba_{J^{\T}}^*\ba_{I}}e^{-\ba^*\tA\ba+\ba^{\ast}\tilu^\dag \bet+\bet^{\ast}\tilu \ba}|_{\ba=\ba^{\ast
}=0}$ is only related to the coefficient of $\alpha_d^{i_d}\cdots\alpha_1^{i_1}\alpha_1^{*j_1}\cdots\alpha_d^{*j_d}$ in the polynomial expansion of $e^{-\ba^*\tA\ba+\ba^{\ast}\tilu^\dag \bet+\bet^{\ast}\tilu \ba}$. Note 
\begin{align}
\label{eq_expansion2}
e^{-\ba^*\tA\ba+\ba^{\ast}\tilu^\dag \bet+\bet^{\ast}\tilu \ba 
}=\sum_{k_1,k_2,k_3=0}^{\infty }\frac{( -\ba^*\tA\ba )^{k_1}}{k_1!}
\frac{( \bet^{\ast}\tilu \ba) ^{k_2}}{k_2!}
\frac{( \ba^{\ast}\tilu^\dag \bet) ^{k_3}}{k_3!}\,,
\end{align}
the terms with factor $\alpha_d^{i_d}\cdots\alpha_1^{i_1} \alpha_1^{\ast
j_1} \cdots \alpha_d^{*j_d}$ can be obtained through
that 
\begin{enumerate}[(i)]    
\item $\frac{(-\ba^*\tA\ba) ^{i_1^{\prime
}+\cdots +i_d'}}{( i_1'+\cdots +i_d^{\prime
}) !}$ contributes to $\alpha_d^{i'_d}\cdots\alpha_1^{i'_1} \alpha_1^{\ast
j'_1} \cdots \alpha_d^{*j'_d}$,

\item $\frac{( \bet^{\ast}\tilu \ba) ^{\overline{i_1'}%
+\cdots +\overline{i_d'}}}{( \overline{i_1'}%
+\cdots +\overline{i_d'}) !}$ contributes to $\alpha_d^{\overline{i_d'}}\cdots \alpha_1^{\overline{i_1'}}$,

\item $\frac{( \ba^{\ast}\tilu^\dag \bet) ^{\overline{j_1'}%
+\cdots +\overline{j_d'}}}{( \overline{j_1'}%
+\cdots +\overline{j_d'}) !}$ contributes to $ \alpha_1^{*\overline{j_1'}} \cdots \alpha
_d^{*\overline{j_d'}}$, 
\end{enumerate}
where $i'_1,\cdots,i'_d, j'_1,\cdots,j'_d, \overline{i'_1},\cdots,\overline{i'_d}, \overline{j'_1},\cdots,\overline{j'_d}$ satisfy the constraint 
\begin{subequations}
\label{eq_app_con_fer}
\begin{align}
&i'_1,\cdots,i'_d, j'_1,\cdots,j'_d, \overline{i'_1},\cdots,\overline{i'_d}, \overline{j'_1},\cdots,\overline{j'_d}\in \{0,1\}\,; \\
&\overline{i_1'
}=i_1-i_1',\quad \cdots,\quad  \overline{i_d'}=i_d-i_d'\,; \\
&\overline{j_1'}=j_1-j_1', \quad \cdots, \quad  \overline{j_d'}=j_d-j_d'\,;\\
&i_1'+\cdots
+i_d'=j_1'+\cdots +j_d'\,.  
\end{align}
\end{subequations}
 Note that 
\begin{enumerate}[(i)]
    \item the coefficient of $\alpha_d^{i'_d}\cdots\alpha_1^{i'_1} \alpha_1^{\ast
j'_1} \cdots \alpha_d^{*j'_d}$ in $\frac{( -\ba^*\tA\ba 
) ^{i_1'+\cdots +i_d'}}{( i_1^{\prime
}+\cdots +i_d') !}$ is $\det(\tA_{J',I'})$;
\item the coefficient of $\alpha_d^{\overline{i_d'}} \cdots \alpha_1^{\overline{i_1'}}$ in $\frac{( \bet^{\ast}\tilu \ba) ^{\overline{i_1'}+\cdots +\overline{i_d^{\prime
}}}}{( \overline{i_1'}+\cdots +\overline{i_d'}%
) !}$ is ${(\bet^{\ast}\tilu)_{\overline{I'}}}$;
\item the coefficient of $\alpha_1^{*\overline{j_1'}} \cdots \alpha_d^{*\overline{j_d'}}$ in $\frac{(
\ba^{\ast}\tilu^\dag \bet) ^{\overline{j_1'}+\cdots +\overline{j_d'}}}{( \overline{j_1'}+\cdots +\overline{j_d'}) !}$ is ${(\tilu^\dag \bet)_{\overline{J'}^{\rm{T}}}}$;
\end{enumerate}
and 
\begin{align}
\alpha_d^{i_d}\cdots\alpha_1^{i_1} \alpha_1^{\ast
j_1} \cdots \alpha_d^{*j_d} =& \delta^I_{I'\overline{I'}}\delta^J_{J'\overline{J'}} \alpha_d^{i'_d}\cdots\alpha_1^{i'_1} \alpha_1^{\ast
j'_1} \cdots \alpha_d^{*j'_d} \alpha_d^{\overline{i_d'}}
\cdots
\alpha_1^{\overline{i_1'}}
\alpha_1^{*\overline{j_1'}}
\cdots
 \alpha_d^{*\overline{j_d'}}
\,,
\end{align}
therefore, the coefficient of $\alpha_d^{i_d}\cdots\alpha_1^{i_1} \alpha_1^{\ast
j_1} \cdots \alpha_d^{*j_d}$ in $\frac{
(-\ba^*\tA\ba) ^{i_1'+\cdots
+i_d'}}{( i_1'+\cdots +i_d') !}
\frac{( \ba^{\ast}\tilu^\dag \bet) ^{\overline{j_1'}+\cdots +
\overline{j_d'}}}{( \overline{j_1'}+\cdots +
\overline{j_d'}) !}\frac{( \bet^{\ast}\tilu \ba) ^{
\overline{i_1'}+\cdots +\overline{i_d'}}}{( 
\overline{i_1'}+\cdots +\overline{i_d'}) !}$ is 
$\delta^I_{I'\overline{I'}}\delta^J_{J'\overline{J'}}  \det(\tA_{J',I'}) {(\bet^{\ast}\tilu)_{\overline{I'}}}  {(\tilu^\dag \bet)_{\overline{J'}^{\rm{T}}}}$.
The total coefficient of $\alpha_d^{i_d}\cdots\alpha_1^{i_1} \alpha_1^{\ast
j_1} \cdots \alpha_d^{*j_d}$ in Eq.~\eqref{eq_expansion2} is
the summation of all the possible terms with $i_{( \cdot ) }'$'s $j_{(
\cdot ) }'$'s satisfying Eq.~\eqref{eq_app_con_fer}, which is denoted as
\begin{align}
\sum_{i_1',\cdots ,i_d',j_1',\cdots
,j_d'} \delta^I_{I'\overline{I'}}\delta^J_{J'\overline{J'}}  \det(\tA_{J',I'}) {(\bet^{\ast}\tilu)_{\overline{I'}}}  {(\tilu^\dag \bet)_{\overline{J'}^{\rm{T}}}}\,.
\end{align}
Also note that for Grassmannian variables
\begin{align}
\partial _{\ba_{J^{\T}}^*\ba_{I}}( \alpha_d^{i_d}\cdots\alpha_1^{i_1} \alpha_1^{\ast
j_1} \cdots \alpha_d^{*j_d}) =1 \,,
\end{align}
therefore
\begin{align}
\eval{\partial _{\ba_{J^{\T}}^*\ba_{I}}e^{-\ba^*\tA%
\ba+\ba^{\ast}\tilu^\dag \bet+\bet^{\ast}\tilu \ba}}_{\ba=\ba^*=0}
=\sum_{i_1',\cdots,i_d',j_1',\cdots,j_d'} \delta^I_{I'\overline{I'}}\delta^J_{J'\overline{J'}} \det(\tA_{J',I'})(\bet^{\ast}\tilu)_{\overline{I'}}(\tilu^\dag\bet)_{\overline{J'}^{\rm{T}}}\,.
\end{align}
 We can express the result in a new way, i.e., as the summation over all the possible ways of obtaining $I'$'s and $J'$'s from $I$ and $J$, respectively. Because the way obtaining $I'$ and $J'$ from $I$ and $J$ in the fermionic case is unique, $\sum_{i_1',\cdots,i_d',j_1',\cdots,j_d'}$ can be directly replaced by $\sum_{I',J'}$. Therefore, the final result can be formulated as
\begin{align}
\int\d\mu(\bz)\d\mu(\bz') \bz^*_I\bz'_{J^{\T}} e^{- {\bz}^{\prime\ast}\tA\bz+{\bz}^{\prime\ast}\tilu^\dag{\bet}+{\bet}^{\ast}\tilu \bz}=\sum_{I',J'} \delta^I_{I'\overline{I'}}\delta^J_{J'\overline{J'}} \det( \tA_{J^{\prime
},I'}) (\bet^{\ast}\tilu)_{\overline{I'}}(\tilu^\dag\bet)_{\overline{J'}^{\rm{T}}}\,.
\end{align}

\section{Asymptotic form of 
\texorpdfstring{$\v(t,t)$}{p} as $t$ approaches infinity}
\label{app_b}

Except for the formula in Ref.~\cite{ZLX12}, the elements of $\v(t,t)$ can also be expressed in terms of the system-environment Green function, reading
\begin{align}
v_{ii'}(t,t)
=\sum_k u_{ik}(t) f(\e_k) u^\dag_{ki'}(t)\,, 
\end{align}
where $u_{ik}(t)=\langle[a_i(t),b_k^\dag(0)]_{\mp}\rangle$ and $u^\dag_{ki'}(t)=[u_{i'k}(t)]^*$. In order to obtain $\v(t,t)$ in the long-time limit, we need to find the asymptotic behavior of  $u_{ik}(t)$. 

Note that
\begin{align}
u_{ik}(t)&=\sum_j \int_0^t \d \t u_{ij}(t-\t) {V}_{jk} e^{-\ri {\e}_k \t}=\ri \sum_j \int_{-\infty}^{\infty} \frac{\d \e}{2 \pi} U_{ij}(\e+\ri 0^+) {V}_{jk} e^{- \ri \e t} \int_0^t \d \t  e^{ \ri (\e-{\e}_k) \t}\,,
\end{align}
and
\begin{align}
\int_0^\infty \d \t  e^{ \ri (\e-{\e}_k) \t}=\frac{\ri}{\e-{\e}_k+\ri 0^+}\,,
\end{align}
$\lim_{t\ra \infty}u_{ik}(t)$ can be simplified to
\begin{align}
\lim_{t\ra \infty}u_{ik}(t)&=- \sum_j  \int_{-\infty}^{\infty} \frac{\d \e}{2 \pi} \frac{U_{ij}(\e+\ri 0^+) {V}_{jk} }{\e-{\e}_k+\ri 0^+} e^{- \ri \e t}\,. \label{eq_uik2}
\end{align}
When there are no localized modes in the total system, there is no singularity in $U_{ij}(\e+\ri 0^+)$ and there is only a pole located at $\e_k-\ri 0$ in the integrated function of Eq.~\eqref{eq_uik2}. Using the contour integral, one obtains
\begin{align}
\lim_{t\ra \infty}u_{ik}(t)
&=-\sum_{j}U_{ij}({\e}_k+\ri 0^+) {V}_{jk} e^{- \ri \e_k t}\,.
\end{align}

$v_{ii'}(t,t)$ in the long-time limit therefore reads
\begin{align}
v_{ii'}(t,t)
&=\int \frac{\d \e}{2 \pi} f(\e) \sum_{j,j'} U_{ij}(\e+\ri 0^+)  e^{- \ri \e t}  J_{jj'} (\e) U^\dag_{j'i'}(\e+\ri 0^+)  e^{\ri \e t}\,. 
\end{align}
In terms of the matrix representation, it can be expressed as
\begin{align}\label{eq_vttappc}
\lim_{t\ra \infty}\v(t,t)=\int \frac{\d \e}{2 \pi} f(\e) \U(\e+\ri 0^+)  \J (\e) \U^\dag(\e+\ri 0^+)\,. 
\end{align}
Following Eq.~\eqref{eq_D2}, it can be further simplified to
\begin{align}
\lim_{t\ra \infty}\v(t,t)
=\int \frac{\d \e}{2 \pi} f(\e)\D(\e)\,.
\end{align}
This is the standard equilibrium fluctuation-dissipation theorem.

\section{Equilibrium state of the system}
\label{app_c}

If the total system possesses no localized modes, the open system would finally reach an equilibrium state. Following the expression of the reduced density matrix in Eq.~\eqref{eq_rhot} and the asymptotic expression of $\u(t)$ and $\v(t,t)$ in Eq.~\eqref{eq_uv}, when $t$ approaches infinity, only the terms with $\overline{I'}=\overline{J'}=0$ would survive, i.e.,
\begin{align}
\rho(\infty)=\sum_{IJ}{\rho_{IJ}(0)} \frac{|\tA_{J,I}|_\pm \rho^{{\rm{th}}}(\infty)} {\sqrt{i_1 !\cdots i_d ! j_1 ! \cdots j_d !}} \,.\label{eq_rho_inf_1}
\end{align}
Because $\tA$ reduces to the identity matrix $\bm{\rm{I}}$ as $\u$ vanishes, the matrix $\tA_{J,I}$ would approach to a block-diagonal matrix in the form
\begin{equation}
\tA_{J,I}= 
\mqty(\dmat{(\tA_{J,I})_1,(\tA_{J,I})_2,\ddots,(\tA_{J,I})_d}) 
\end{equation}
where 
\begin{equation}
(\tA_{J,I})_n=\begin{pmatrix}
1 & \cdots & 1  \\
\vdots & \ddots & \vdots \\
1 & \cdots & 1 
\end{pmatrix}_{j_n \times i_n} \,.
\end{equation}
The \emph{permanent} and \emph{determinant} of $\tA_{J,I}$ then share a common expression, reading
\begin{equation}
|\tA_{J,I}|_\pm=
\left\{
\begin{array}{lr}
0  &\quad (I \neq J)\,,\\ 
i_1 !\cdots i_d !  &\quad (I=J) \,. 
\end{array}
\right.
\end{equation}   
Following Eq.~\eqref{eq_rho_inf_1}, $\rho(\infty)$ can be simplified to
\begin{align}
\rho(\infty)=\sum_{I=J}{\rho_{IJ}(0)} \rho^{{\rm{th}}}(\infty)\,.\label{eq_rho_inf_2}
\end{align}
With the normalization condition that $\sum_{I=J}{\rho_{IJ}(0)}=1$, it can be finally written as
\begin{align}
\rho(\infty)=\rho^{{\rm{th}}}(\infty)=\frac{e^{ \a^{\dag}\ln(\frac{\v(\infty,\infty)}{1\pm \v(\infty,\infty)}) {\a}}}{\vert 1\pm \v(\infty,\infty)\vert^{\pm 1}}=\frac{e^{ \a^{\dag}\ln(\frac{\overline{\bm{n}}}{1\pm \overline{\bm{n}}}) {\a}}}{\vert 1\pm \overline{\bm{n}}\vert^{\pm 1}} \,.
\end{align}

\section{Asymptotic form of \texorpdfstring{$\D(\e)$}{p} as the coupling strength vanishes}
\label{app_d}


When $\J(\e)\ra \bm{0}$ and $\bm{\Delta}(\e) \ra \bm{0}$, the spectrum
\begin{align}\label{eq_D2}
\bm{D}(\e)
=\frac{1}{\e-\bm{\e}_\S -\bm{\Delta}(\e) +\ri \frac{\bm{J}(\e)}{2}} \J(\e) \frac{1}{\e-\bm{\e}_\S -\bm{\Delta}(\e) -\ri \frac{\bm{J}(\e)}{2}}\,,
\end{align}
is vanishing for the values of $\e$ that the matrices $\e-\bm{\e}_\S -\bm{\Delta}(\e) \pm\ri \frac{\bm{J}(\e)}{2}$ are invertible. Because $\J(\e)\ra \bm{0}$ and $\bm{\Delta}(\e) \ra \bm{0}$, the condition can be simplified as that $\e-\bm{\e}_\S$ is invertible. Therefore, $\bm{D}(\e)$ is nonvanishing only for $\e$ equaling an eigenvalue $\e_\lambda$ of $\bm{\e}_\S$. So in order to grasp the asymptotic behavior of $\bm{D}(\e)$, one only needs to analyze the behavior of $\bm{D}(\e)$ around $\e=\e_\lambda$.

Consider the behavior of $\frac{1}{\e-\bm{\e}_\S -\bm{\Delta}(\e) \pm \ri \frac{\bm{J}(\e)}{2}}$ for $\e\approx\e_\lambda$. 
Denote the eigenspace of $\bm{\e}_\S$ corresponding to the eigenvalue $\e_\lambda$ as $\mathbb{H}_\lambda$, 
then the matrix ${\e-\bm{\e}_\S -\bm{\Delta}(\e) \pm \ri \frac{\bm{J}(\e)}{2}}$ can be written in blocks, reading
\begin{align}
\e-\bm{\e}_\S -\bm{\Delta}(\e) \pm  \frac{\ri}{2} \bm{J}(\e)
=\begin{pmatrix}
(\e-\e_\lambda)\mathbb{I}_\lambda -\bm{\Delta}_{\lambda\lambda}(\e) \pm  \frac{\ri}{2} \bm{J}_{\lambda\lambda}(\e)
&  \l[-\bm{\Delta}(\e) \pm  \frac{\ri}{2} \bm{J}(\e) \r]_{\lambda\lambda^\perp}
\\
\l[-\bm{\Delta}(\e) \pm  \frac{\ri}{2} \bm{J}(\e) \r]_{\lambda^\perp \lambda}
& \l[\e-\bm{\e}_\S -\bm{\Delta}(\e) \pm  \frac{\ri}{2} \bm{J}(\e)\r]_{\lambda^\perp\lambda^\perp}
\end{pmatrix}\,,
\end{align}
where the subscripts $\lambda$ and $\lambda^{\perp}$ are corresponding to the space $\mathbb{H}_\lambda$ and its orthogonal complement, respectively. 
Because $\J(\e)\approx\bm{0}$ and $\bm{\Delta}(\e) \approx \bm{0}$, the inverse of $\e-\bm{\e}_\S -\bm{\Delta}(\e) \pm  \frac{\ri}{2} \bm{J}(\e)$ is approximately
\begin{align}
\frac{1}{\e-\bm{\e}_\S -\bm{\Delta}(\e) \pm  \frac{\ri}{2} \bm{J}(\e)}
\approx\begin{pmatrix}
\frac{1}{(\e-\e_\lambda)\mathbb{I}_\lambda -\bm{\Delta}_{\lambda\lambda}(\e) \pm  \frac{\ri}{2} \bm{J}_{\lambda\lambda}(\e)} &  \bm{0}  \\
\bm{0} & \frac{1}{\e-(\bm{\e}_\S)_{\lambda^\perp\lambda^\perp}} 
\end{pmatrix}\,,\label{eq_aprx}
\end{align}
where $\mathbb{I}_\lambda$ is the identity in $\mathbb{H}_\lambda$. (In this equation, we have kept the term $(\e-\e_\lambda)\mathbb{I}_\lambda$ for $\e-\e_\lambda$ could also be small.) Following  Eqs.~\eqref{eq_D2} and~\eqref{eq_aprx}, and using the properties that $\J(\e)\approx\bm{0}$ and $\bm{\Delta}(\e) \approx \bm{0}$, the expression of $\bm{D}(\e)$ around $\e\approx \e_\lambda$ can be approximately written as 
\begin{align}
\bm{D}(\e)
\approx\frac{1}{(\e-\e_\lambda)\mathbb{I}_\lambda -\bm{\Delta}_{\lambda\lambda}(\e) + \frac{\ri}{2} \bm{J}_{\lambda\lambda}(\e)}
\bm{J}_{\lambda\lambda}(\e)
\frac{1}{(\e-\e_\lambda)\mathbb{I}_\lambda -\bm{\Delta}_{\lambda\lambda}(\e) - \frac{\ri}{2} \bm{J}_{\lambda\lambda}(\e)}\,.
\end{align}
When $\bm{J}(\e)$ and $\bm{\Delta}(\e)$ approach to $\bm{0}$, the real part of $(\e-\e_\lambda)\mathbb{I}_\lambda -\bm{\Delta}_{\lambda\lambda}(\e) \pm \frac{\ri}{2} \bm{J}_{\lambda\lambda}(\e)$ is dominant by $(\e-\e_\lambda)\mathbb{I}_\lambda$, so the above equation can be further simplified to 
\begin{align}
\bm{D}(\e)
\approx\frac{\bm{J}_{\lambda\lambda}(\e)}{(\e-\e_\lambda)^2\mathbb{I}_\lambda + [\bm{J}_{\lambda\lambda}(\e)]^2/4}
\,.
\end{align}
After expressing the right hand side of the equation in the eigenbasis of $\bm{J}_{\lambda\lambda}(\e)$, one can easily find that the diagonal elements all approach to $2\pi \delta(\e-\e_\lambda)$ as $\bm{J}(\e)$ vanishes. Consequently, for $\e$ near $\e_\lambda$, 
\begin{align}
\bm{D}(\e)
\ra 2\pi \delta(\e-\e_\lambda) \mathbb{I}_\lambda
\,.
\end{align}
For every $\lambda$, such conclusion is always true. Therefore, for all $\e$, 
\begin{align}
\bm{D}(\e)
\ra 2\pi \sum_\lambda \delta(\e-\e_\lambda) \mathbb{I}_\lambda=2\pi \delta (\e \rm{\bm{I}} -\bm{\e}_\S)\,. 
\end{align}

\end{widetext}    
\end{appendix}    

\bibliographystyle{apsrev4-1}
\bibliography{references}

\begin{thebibliography}{62}%
\makeatletter
\providecommand \@ifxundefined [1]{%
 \@ifx{#1\undefined}
}%
\providecommand \@ifnum [1]{%
 \ifnum #1\expandafter \@firstoftwo
 \else \expandafter \@secondoftwo
 \fi
}%
\providecommand \@ifx [1]{%
 \ifx #1\expandafter \@firstoftwo
 \else \expandafter \@secondoftwo
 \fi
}%
\providecommand \natexlab [1]{#1}%
\providecommand \enquote  [1]{``#1''}%
\providecommand \bibnamefont  [1]{#1}%
\providecommand \bibfnamefont [1]{#1}%
\providecommand \citenamefont [1]{#1}%
\providecommand \href@noop [0]{\@secondoftwo}%
\providecommand \href [0]{\begingroup \@sanitize@url \@href}%
\providecommand \@href[1]{\@@startlink{#1}\@@href}%
\providecommand \@@href[1]{\endgroup#1\@@endlink}%
\providecommand \@sanitize@url [0]{\catcode `\\12\catcode `\$12\catcode
  `\&12\catcode `\#12\catcode `\^12\catcode `\_12\catcode `\%12\relax}%
\providecommand \@@startlink[1]{}%
\providecommand \@@endlink[0]{}%
\providecommand \url  [0]{\begingroup\@sanitize@url \@url }%
\providecommand \@url [1]{\endgroup\@href {#1}{\urlprefix }}%
\providecommand \urlprefix  [0]{URL }%
\providecommand \Eprint [0]{\href }%
\providecommand \doibase [0]{http://dx.doi.org/}%
\providecommand \selectlanguage [0]{\@gobble}%
\providecommand \bibinfo  [0]{\@secondoftwo}%
\providecommand \bibfield  [0]{\@secondoftwo}%
\providecommand \translation [1]{[#1]}%
\providecommand \BibitemOpen [0]{}%
\providecommand \bibitemStop [0]{}%
\providecommand \bibitemNoStop [0]{.\EOS\space}%
\providecommand \EOS [0]{\spacefactor3000\relax}%
\providecommand \BibitemShut  [1]{\csname bibitem#1\endcsname}%
\let\auto@bib@innerbib\@empty
\bibitem [{\citenamefont {Feynman}\ and\ \citenamefont {Vernon}(1963)}]{FV63}%
  \BibitemOpen
  \bibfield  {author} {\bibinfo {author} {\bibfnamefont {R.}~\bibnamefont
  {Feynman}}\ and\ \bibinfo {author} {\bibfnamefont {F.}~\bibnamefont
  {Vernon}},\ }\href
  {http://www.sciencedirect.com/science/article/pii/000349166390068X}
  {\bibfield  {journal} {\bibinfo  {journal} {Ann. Phys.}\ }\textbf {\bibinfo
  {volume} {24}},\ \bibinfo {pages} {118 } (\bibinfo {year}
  {1963})}\BibitemShut {NoStop}%
\bibitem [{\citenamefont {Schwinger}(1961)}]{S61}%
  \BibitemOpen
  \bibfield  {author} {\bibinfo {author} {\bibfnamefont {J.}~\bibnamefont
  {Schwinger}},\ }\href {https://aip.scitation.org/doi/10.1063/1.1703727}
  {\bibfield  {journal} {\bibinfo  {journal} {J. Math. Phys.}\ }\textbf
  {\bibinfo {volume} {2}},\ \bibinfo {pages} {407} (\bibinfo {year}
  {1961})}\BibitemShut {NoStop}%
\bibitem [{\citenamefont {Zwanzig}(1960)}]{Z60}%
  \BibitemOpen
  \bibfield  {author} {\bibinfo {author} {\bibfnamefont {R.}~\bibnamefont
  {Zwanzig}},\ }\href {https://aip.scitation.org/doi/abs/10.1063/1.1731409}
  {\bibfield  {journal} {\bibinfo  {journal} {Int. J. Chem. Phys.}\ }\textbf
  {\bibinfo {volume} {33}},\ \bibinfo {pages} {1338} (\bibinfo {year}
  {1960})}\BibitemShut {NoStop}%
\bibitem [{\citenamefont {Nakajima}(1958)}]{N58}%
  \BibitemOpen
  \bibfield  {author} {\bibinfo {author} {\bibfnamefont {S.}~\bibnamefont
  {Nakajima}},\ }\href {https://academic.oup.com/ptp/article/20/6/948/1930693}
  {\bibfield  {journal} {\bibinfo  {journal} {Prog. Theor. Phys.}\ }\textbf
  {\bibinfo {volume} {20}},\ \bibinfo {pages} {948} (\bibinfo {year}
  {1958})}\BibitemShut {NoStop}%
\bibitem [{\citenamefont {Breuer}\ \emph {et~al.}(2002)\citenamefont {Breuer},
  \citenamefont {Petruccione} \emph {et~al.}}]{BP02}%
  \BibitemOpen
  \bibfield  {author} {\bibinfo {author} {\bibfnamefont {H.-P.}\ \bibnamefont
  {Breuer}}, \bibinfo {author} {\bibfnamefont {F.}~\bibnamefont {Petruccione}},
   \emph {et~al.},\ }\href@noop {} {\emph {\bibinfo {title} {The Theory of Open
  Quantum Systems}}}\ (\bibinfo  {publisher} {Oxford University Press on
  Demand},\ \bibinfo {year} {2002})\BibitemShut {NoStop}%
\bibitem [{\citenamefont {Weiss}(2012)}]{W12}%
  \BibitemOpen
  \bibfield  {author} {\bibinfo {author} {\bibfnamefont {U.}~\bibnamefont
  {Weiss}},\ }\href@noop {} {\emph {\bibinfo {title} {Quantum Dissipative
  Systems}}},\ Vol.~\bibinfo {volume} {13}\ (\bibinfo  {publisher} {World
  scientific},\ \bibinfo {year} {2012})\BibitemShut {NoStop}%
\bibitem [{\citenamefont {Gardiner}\ and\ \citenamefont {Zoller}(2004)}]{GP04}%
  \BibitemOpen
  \bibfield  {author} {\bibinfo {author} {\bibfnamefont {C.}~\bibnamefont
  {Gardiner}}\ and\ \bibinfo {author} {\bibfnamefont {P.}~\bibnamefont
  {Zoller}},\ }\href@noop {} {\emph {\bibinfo {title} {Quantum Noise: A
  Handbook of Markovian and Non-Markovian Quantum Stochastic Methods with
  Applications to Quantum Optics}}},\ Vol.~\bibinfo {volume} {56}\ (\bibinfo
  {publisher} {Springer Science \& Business Media},\ \bibinfo {year}
  {2004})\BibitemShut {NoStop}%
\bibitem [{\citenamefont {Wiseman}\ and\ \citenamefont {Milburn}(2009)}]{WM09}%
  \BibitemOpen
  \bibfield  {author} {\bibinfo {author} {\bibfnamefont {H.~M.}\ \bibnamefont
  {Wiseman}}\ and\ \bibinfo {author} {\bibfnamefont {G.~J.}\ \bibnamefont
  {Milburn}},\ }\href@noop {} {\emph {\bibinfo {title} {Quantum Measurement and
  Control}}}\ (\bibinfo  {publisher} {Cambridge university press},\ \bibinfo
  {year} {2009})\BibitemShut {NoStop}%
\bibitem [{\citenamefont {Nielsen}\ and\ \citenamefont {Chuang}(2002)}]{NC02}%
  \BibitemOpen
  \bibfield  {author} {\bibinfo {author} {\bibfnamefont {M.~A.}\ \bibnamefont
  {Nielsen}}\ and\ \bibinfo {author} {\bibfnamefont {I.}~\bibnamefont
  {Chuang}},\ }\href@noop {} {\enquote {\bibinfo {title} {Quantum computation
  and quantum information},}\ } (\bibinfo {year} {2002})\BibitemShut {NoStop}%
\bibitem [{\citenamefont {Haug}\ and\ \citenamefont {Jauho}(2008)}]{HJ08}%
  \BibitemOpen
  \bibfield  {author} {\bibinfo {author} {\bibfnamefont {H.}~\bibnamefont
  {Haug}}\ and\ \bibinfo {author} {\bibfnamefont {A.-P.}\ \bibnamefont
  {Jauho}},\ }\href@noop {} {\emph {\bibinfo {title} {Quantum Kinetics in
  Transport and Optics of Semiconductors}}},\ Vol.~\bibinfo {volume} {2}\
  (\bibinfo  {publisher} {Springer},\ \bibinfo {year} {2008})\BibitemShut
  {NoStop}%
\bibitem [{\citenamefont {Yang}\ and\ \citenamefont {Zhang}(2016)}]{YZ16}%
  \BibitemOpen
  \bibfield  {author} {\bibinfo {author} {\bibfnamefont {P.-Y.}\ \bibnamefont
  {Yang}}\ and\ \bibinfo {author} {\bibfnamefont {W.-M.}\ \bibnamefont
  {Zhang}},\ }\href {https://doi.org/10.1007/s11467-016-0640-z} {\bibfield
  {journal} {\bibinfo  {journal} {Front. Phys.}\ }\textbf {\bibinfo {volume}
  {12}},\ \bibinfo {pages} {127204} (\bibinfo {year} {2016})}\BibitemShut
  {NoStop}%
\bibitem [{\citenamefont {Wickenhauser}\ \emph {et~al.}(2005)\citenamefont
  {Wickenhauser}, \citenamefont {Burgd\"orfer}, \citenamefont {Krausz},\ and\
  \citenamefont {Drescher}}]{WBK05}%
  \BibitemOpen
  \bibfield  {author} {\bibinfo {author} {\bibfnamefont {M.}~\bibnamefont
  {Wickenhauser}}, \bibinfo {author} {\bibfnamefont {J.}~\bibnamefont
  {Burgd\"orfer}}, \bibinfo {author} {\bibfnamefont {F.}~\bibnamefont
  {Krausz}}, \ and\ \bibinfo {author} {\bibfnamefont {M.}~\bibnamefont
  {Drescher}},\ }\href {https://link.aps.org/doi/10.1103/PhysRevLett.94.023002}
  {\bibfield  {journal} {\bibinfo  {journal} {Phys. Rev. Lett.}\ }\textbf
  {\bibinfo {volume} {94}},\ \bibinfo {pages} {023002} (\bibinfo {year}
  {2005})}\BibitemShut {NoStop}%
\bibitem [{\citenamefont {Kaldun}\ \emph {et~al.}(2016)\citenamefont {Kaldun},
  \citenamefont {Bl{\"a}ttermann}, \citenamefont {Stoo{\ss}}, \citenamefont
  {Donsa}, \citenamefont {Wei}, \citenamefont {Pazourek}, \citenamefont
  {Nagele}, \citenamefont {Ott}, \citenamefont {Lin}, \citenamefont
  {Burgd{\"o}rfer} \emph {et~al.}}]{KBS16}%
  \BibitemOpen
  \bibfield  {author} {\bibinfo {author} {\bibfnamefont {A.}~\bibnamefont
  {Kaldun}}, \bibinfo {author} {\bibfnamefont {A.}~\bibnamefont
  {Bl{\"a}ttermann}}, \bibinfo {author} {\bibfnamefont {V.}~\bibnamefont
  {Stoo{\ss}}}, \bibinfo {author} {\bibfnamefont {S.}~\bibnamefont {Donsa}},
  \bibinfo {author} {\bibfnamefont {H.}~\bibnamefont {Wei}}, \bibinfo {author}
  {\bibfnamefont {R.}~\bibnamefont {Pazourek}}, \bibinfo {author}
  {\bibfnamefont {S.}~\bibnamefont {Nagele}}, \bibinfo {author} {\bibfnamefont
  {C.}~\bibnamefont {Ott}}, \bibinfo {author} {\bibfnamefont {C.}~\bibnamefont
  {Lin}}, \bibinfo {author} {\bibfnamefont {J.}~\bibnamefont {Burgd{\"o}rfer}},
   \emph {et~al.},\ }\href
  {https://science.sciencemag.org/content/354/6313/738.abstract} {\bibfield
  {journal} {\bibinfo  {journal} {Science}\ }\textbf {\bibinfo {volume}
  {354}},\ \bibinfo {pages} {738} (\bibinfo {year} {2016})}\BibitemShut
  {NoStop}%
\bibitem [{\citenamefont {Caldeira}\ and\ \citenamefont
  {Leggett}(1983)}]{CL83}%
  \BibitemOpen
  \bibfield  {author} {\bibinfo {author} {\bibfnamefont {A.}~\bibnamefont
  {Caldeira}}\ and\ \bibinfo {author} {\bibfnamefont {A.}~\bibnamefont
  {Leggett}},\ }\href
  {http://www.sciencedirect.com/science/article/pii/0378437183900134}
  {\bibfield  {journal} {\bibinfo  {journal} {Physica A}\ }\textbf {\bibinfo
  {volume} {121}},\ \bibinfo {pages} {587 } (\bibinfo {year}
  {1983})}\BibitemShut {NoStop}%
\bibitem [{\citenamefont {Hu}\ \emph {et~al.}(1992)\citenamefont {Hu},
  \citenamefont {Paz},\ and\ \citenamefont {Zhang}}]{HPZ92}%
  \BibitemOpen
  \bibfield  {author} {\bibinfo {author} {\bibfnamefont {B.~L.}\ \bibnamefont
  {Hu}}, \bibinfo {author} {\bibfnamefont {J.~P.}\ \bibnamefont {Paz}}, \ and\
  \bibinfo {author} {\bibfnamefont {Y.}~\bibnamefont {Zhang}},\ }\href
  {https://link.aps.org/doi/10.1103/PhysRevD.45.2843} {\bibfield  {journal}
  {\bibinfo  {journal} {Phys. Rev. D}\ }\textbf {\bibinfo {volume} {45}},\
  \bibinfo {pages} {2843} (\bibinfo {year} {1992})}\BibitemShut {NoStop}%
\bibitem [{\citenamefont {Zhang}\ \emph {et~al.}(2012)\citenamefont {Zhang},
  \citenamefont {Lo}, \citenamefont {Xiong}, \citenamefont {Tu},\ and\
  \citenamefont {Nori}}]{ZLX12}%
  \BibitemOpen
  \bibfield  {author} {\bibinfo {author} {\bibfnamefont {W.-M.}\ \bibnamefont
  {Zhang}}, \bibinfo {author} {\bibfnamefont {P.-Y.}\ \bibnamefont {Lo}},
  \bibinfo {author} {\bibfnamefont {H.-N.}\ \bibnamefont {Xiong}}, \bibinfo
  {author} {\bibfnamefont {M.~W.-Y.}\ \bibnamefont {Tu}}, \ and\ \bibinfo
  {author} {\bibfnamefont {F.}~\bibnamefont {Nori}},\ }\href
  {https://journals.aps.org/prl/abstract/10.1103/PhysRevLett.109.170402}
  {\bibfield  {journal} {\bibinfo  {journal} {Phys. Rev. Lett.}\ }\textbf
  {\bibinfo {volume} {109}},\ \bibinfo {pages} {170402} (\bibinfo {year}
  {2012})}\BibitemShut {NoStop}%
\bibitem [{\citenamefont {Tu}\ and\ \citenamefont {Zhang}(2008)}]{TZ08}%
  \BibitemOpen
  \bibfield  {author} {\bibinfo {author} {\bibfnamefont {M.~W.~Y.}\
  \bibnamefont {Tu}}\ and\ \bibinfo {author} {\bibfnamefont {W.-M.}\
  \bibnamefont {Zhang}},\ }\href
  {https://link.aps.org/doi/10.1103/PhysRevB.78.235311} {\bibfield  {journal}
  {\bibinfo  {journal} {Phys. Rev. B}\ }\textbf {\bibinfo {volume} {78}},\
  \bibinfo {pages} {235311} (\bibinfo {year} {2008})}\BibitemShut {NoStop}%
\bibitem [{\citenamefont {Jin}\ \emph {et~al.}(2010)\citenamefont {Jin},
  \citenamefont {Tu}, \citenamefont {Zhang},\ and\ \citenamefont
  {Yan}}]{JTZY10}%
  \BibitemOpen
  \bibfield  {author} {\bibinfo {author} {\bibfnamefont {J.}~\bibnamefont
  {Jin}}, \bibinfo {author} {\bibfnamefont {M.~W.-Y.}\ \bibnamefont {Tu}},
  \bibinfo {author} {\bibfnamefont {W.-M.}\ \bibnamefont {Zhang}}, \ and\
  \bibinfo {author} {\bibfnamefont {Y.}~\bibnamefont {Yan}},\ }\href
  {https://iopscience.iop.org/article/10.1088/1367-2630/12/8/083013} {\bibfield
   {journal} {\bibinfo  {journal} {New J. Phys.}\ }\textbf {\bibinfo {volume}
  {12}},\ \bibinfo {pages} {083013} (\bibinfo {year} {2010})}\BibitemShut
  {NoStop}%
\bibitem [{\citenamefont {Lei}\ and\ \citenamefont {Zhang}(2012)}]{LZ12}%
  \BibitemOpen
  \bibfield  {author} {\bibinfo {author} {\bibfnamefont {C.~U.}\ \bibnamefont
  {Lei}}\ and\ \bibinfo {author} {\bibfnamefont {W.-M.}\ \bibnamefont
  {Zhang}},\ }\href
  {https://www.sciencedirect.com/science/article/pii/S0003491612000231}
  {\bibfield  {journal} {\bibinfo  {journal} {Ann. Phys.}\ }\textbf {\bibinfo
  {volume} {327}},\ \bibinfo {pages} {1408} (\bibinfo {year}
  {2012})}\BibitemShut {NoStop}%
\bibitem [{\citenamefont {Anderson}(1961)}]{A61}%
  \BibitemOpen
  \bibfield  {author} {\bibinfo {author} {\bibfnamefont {P.~W.}\ \bibnamefont
  {Anderson}},\ }\href {https://link.aps.org/doi/10.1103/PhysRev.124.41}
  {\bibfield  {journal} {\bibinfo  {journal} {Phys. Rev.}\ }\textbf {\bibinfo
  {volume} {124}},\ \bibinfo {pages} {41} (\bibinfo {year} {1961})}\BibitemShut
  {NoStop}%
\bibitem [{\citenamefont {Fano}(1961)}]{F61}%
  \BibitemOpen
  \bibfield  {author} {\bibinfo {author} {\bibfnamefont {U.}~\bibnamefont
  {Fano}},\ }\href {https://link.aps.org/doi/10.1103/PhysRev.124.1866}
  {\bibfield  {journal} {\bibinfo  {journal} {Phys. Rev.}\ }\textbf {\bibinfo
  {volume} {124}},\ \bibinfo {pages} {1866} (\bibinfo {year}
  {1961})}\BibitemShut {NoStop}%
\bibitem [{\citenamefont {Miroshnichenko}\ \emph {et~al.}(2010)\citenamefont
  {Miroshnichenko}, \citenamefont {Flach},\ and\ \citenamefont
  {Kivshar}}]{MFK10}%
  \BibitemOpen
  \bibfield  {author} {\bibinfo {author} {\bibfnamefont {A.~E.}\ \bibnamefont
  {Miroshnichenko}}, \bibinfo {author} {\bibfnamefont {S.}~\bibnamefont
  {Flach}}, \ and\ \bibinfo {author} {\bibfnamefont {Y.~S.}\ \bibnamefont
  {Kivshar}},\ }\href {https://link.aps.org/doi/10.1103/RevModPhys.82.2257}
  {\bibfield  {journal} {\bibinfo  {journal} {Rev. Mod. Phys.}\ }\textbf
  {\bibinfo {volume} {82}},\ \bibinfo {pages} {2257} (\bibinfo {year}
  {2010})}\BibitemShut {NoStop}%
\bibitem [{\citenamefont {Anderson}(1958)}]{A58}%
  \BibitemOpen
  \bibfield  {author} {\bibinfo {author} {\bibfnamefont {P.~W.}\ \bibnamefont
  {Anderson}},\ }\href {https://link.aps.org/doi/10.1103/PhysRev.109.1492}
  {\bibfield  {journal} {\bibinfo  {journal} {Phys. Rev.}\ }\textbf {\bibinfo
  {volume} {109}},\ \bibinfo {pages} {1492} (\bibinfo {year}
  {1958})}\BibitemShut {NoStop}%
\bibitem [{\citenamefont {Yablonovitch}(1987)}]{Y87}%
  \BibitemOpen
  \bibfield  {author} {\bibinfo {author} {\bibfnamefont {E.}~\bibnamefont
  {Yablonovitch}},\ }\href
  {https://journals.aps.org/prl/abstract/10.1103/PhysRevLett.58.2059}
  {\bibfield  {journal} {\bibinfo  {journal} {Phys. Rev. Lett.}\ }\textbf
  {\bibinfo {volume} {58}},\ \bibinfo {pages} {2059} (\bibinfo {year}
  {1987})}\BibitemShut {NoStop}%
\bibitem [{\citenamefont {John}(1987)}]{J87}%
  \BibitemOpen
  \bibfield  {author} {\bibinfo {author} {\bibfnamefont {S.}~\bibnamefont
  {John}},\ }\href
  {https://journals.aps.org/prl/abstract/10.1103/PhysRevLett.58.2486}
  {\bibfield  {journal} {\bibinfo  {journal} {Phys. Rev. Lett.}\ }\textbf
  {\bibinfo {volume} {58}},\ \bibinfo {pages} {2486} (\bibinfo {year}
  {1987})}\BibitemShut {NoStop}%
\bibitem [{\citenamefont {Kofman}\ \emph {et~al.}(1994)\citenamefont {Kofman},
  \citenamefont {Kurizki},\ and\ \citenamefont {Sherman}}]{KKS94}%
  \BibitemOpen
  \bibfield  {author} {\bibinfo {author} {\bibfnamefont {A.}~\bibnamefont
  {Kofman}}, \bibinfo {author} {\bibfnamefont {G.}~\bibnamefont {Kurizki}}, \
  and\ \bibinfo {author} {\bibfnamefont {B.}~\bibnamefont {Sherman}},\ }\href
  {https://www.tandfonline.com/doi/abs/10.1080/09500349414550381} {\bibfield
  {journal} {\bibinfo  {journal} {J. Mod. Opt.}\ }\textbf {\bibinfo {volume}
  {41}},\ \bibinfo {pages} {353} (\bibinfo {year} {1994})}\BibitemShut
  {NoStop}%
\bibitem [{\citenamefont {Mahan}(2013)}]{M13}%
  \BibitemOpen
  \bibfield  {author} {\bibinfo {author} {\bibfnamefont {G.~D.}\ \bibnamefont
  {Mahan}},\ }\href@noop {} {\emph {\bibinfo {title} {Many-Particle Physics}}}\
  (\bibinfo  {publisher} {Springer Science \& Business Media},\ \bibinfo {year}
  {2013})\BibitemShut {NoStop}%
\bibitem [{\citenamefont {Yang}\ \emph {et~al.}(2013)\citenamefont {Yang},
  \citenamefont {An}, \citenamefont {Zhang}, \citenamefont {Feng},\ and\
  \citenamefont {Oh}}]{YAZ13}%
  \BibitemOpen
  \bibfield  {author} {\bibinfo {author} {\bibfnamefont {W.~L.}\ \bibnamefont
  {Yang}}, \bibinfo {author} {\bibfnamefont {J.-H.}\ \bibnamefont {An}},
  \bibinfo {author} {\bibfnamefont {C.}~\bibnamefont {Zhang}}, \bibinfo
  {author} {\bibfnamefont {M.}~\bibnamefont {Feng}}, \ and\ \bibinfo {author}
  {\bibfnamefont {C.~H.}\ \bibnamefont {Oh}},\ }\href
  {https://link.aps.org/doi/10.1103/PhysRevA.87.022312} {\bibfield  {journal}
  {\bibinfo  {journal} {Phys. Rev. A}\ }\textbf {\bibinfo {volume} {87}},\
  \bibinfo {pages} {022312} (\bibinfo {year} {2013})}\BibitemShut {NoStop}%
\bibitem [{\citenamefont {Tan}\ \emph {et~al.}(2011)\citenamefont {Tan},
  \citenamefont {Zhang},\ and\ \citenamefont {Li}}]{TZL11}%
  \BibitemOpen
  \bibfield  {author} {\bibinfo {author} {\bibfnamefont {H.-T.}\ \bibnamefont
  {Tan}}, \bibinfo {author} {\bibfnamefont {W.-M.}\ \bibnamefont {Zhang}}, \
  and\ \bibinfo {author} {\bibfnamefont {G.-X.}\ \bibnamefont {Li}},\ }\href
  {https://link.aps.org/doi/10.1103/PhysRevA.83.062310} {\bibfield  {journal}
  {\bibinfo  {journal} {Phys. Rev. A}\ }\textbf {\bibinfo {volume} {83}},\
  \bibinfo {pages} {062310} (\bibinfo {year} {2011})}\BibitemShut {NoStop}%
\bibitem [{\citenamefont {Lin}\ \emph {et~al.}(2016)\citenamefont {Lin},
  \citenamefont {Yang},\ and\ \citenamefont {Zhang}}]{LYZ16}%
  \BibitemOpen
  \bibfield  {author} {\bibinfo {author} {\bibfnamefont {Y.-C.}\ \bibnamefont
  {Lin}}, \bibinfo {author} {\bibfnamefont {P.-Y.}\ \bibnamefont {Yang}}, \
  and\ \bibinfo {author} {\bibfnamefont {W.-M.}\ \bibnamefont {Zhang}},\ }\href
  {https://www.nature.com/articles/srep34804} {\bibfield  {journal} {\bibinfo
  {journal} {Sci. Rep.}\ }\textbf {\bibinfo {volume} {6}},\ \bibinfo {pages}
  {34804} (\bibinfo {year} {2016})}\BibitemShut {NoStop}%
\bibitem [{\citenamefont {An}\ and\ \citenamefont {Zhang}(2007)}]{AZ07}%
  \BibitemOpen
  \bibfield  {author} {\bibinfo {author} {\bibfnamefont {J.-H.}\ \bibnamefont
  {An}}\ and\ \bibinfo {author} {\bibfnamefont {W.-M.}\ \bibnamefont {Zhang}},\
  }\href {https://link.aps.org/doi/10.1103/PhysRevA.76.042127} {\bibfield
  {journal} {\bibinfo  {journal} {Phys. Rev. A}\ }\textbf {\bibinfo {volume}
  {76}},\ \bibinfo {pages} {042127} (\bibinfo {year} {2007})}\BibitemShut
  {NoStop}%
\bibitem [{\citenamefont {Liu}\ \emph {et~al.}(2016)\citenamefont {Liu},
  \citenamefont {Tu},\ and\ \citenamefont {Zhang}}]{LTZ16}%
  \BibitemOpen
  \bibfield  {author} {\bibinfo {author} {\bibfnamefont {J.-H.}\ \bibnamefont
  {Liu}}, \bibinfo {author} {\bibfnamefont {M.~W.-Y.}\ \bibnamefont {Tu}}, \
  and\ \bibinfo {author} {\bibfnamefont {W.-M.}\ \bibnamefont {Zhang}},\ }\href
  {https://link.aps.org/doi/10.1103/PhysRevB.94.045403} {\bibfield  {journal}
  {\bibinfo  {journal} {Phys. Rev. B}\ }\textbf {\bibinfo {volume} {94}},\
  \bibinfo {pages} {045403} (\bibinfo {year} {2016})}\BibitemShut {NoStop}%
\bibitem [{\citenamefont {Yang}\ and\ \citenamefont {Wu}(2014)}]{YW14}%
  \BibitemOpen
  \bibfield  {author} {\bibinfo {author} {\bibfnamefont {M.-J.}\ \bibnamefont
  {Yang}}\ and\ \bibinfo {author} {\bibfnamefont {S.-T.}\ \bibnamefont {Wu}},\
  }\href {https://link.aps.org/doi/10.1103/PhysRevA.89.022301} {\bibfield
  {journal} {\bibinfo  {journal} {Phys. Rev. A}\ }\textbf {\bibinfo {volume}
  {89}},\ \bibinfo {pages} {022301} (\bibinfo {year} {2014})}\BibitemShut
  {NoStop}%
\bibitem [{\citenamefont {Yang}\ and\ \citenamefont {Zhang}(2018)}]{YZ18}%
  \BibitemOpen
  \bibfield  {author} {\bibinfo {author} {\bibfnamefont {P.-Y.}\ \bibnamefont
  {Yang}}\ and\ \bibinfo {author} {\bibfnamefont {W.-M.}\ \bibnamefont
  {Zhang}},\ }\href {https://link.aps.org/doi/10.1103/PhysRevB.97.054301}
  {\bibfield  {journal} {\bibinfo  {journal} {Phys. Rev. B}\ }\textbf {\bibinfo
  {volume} {97}},\ \bibinfo {pages} {054301} (\bibinfo {year}
  {2018})}\BibitemShut {NoStop}%
\bibitem [{\citenamefont {Lo}\ \emph {et~al.}(2015)\citenamefont {Lo},
  \citenamefont {Xiong},\ and\ \citenamefont {Zhang}}]{LXZ15}%
  \BibitemOpen
  \bibfield  {author} {\bibinfo {author} {\bibfnamefont {P.-Y.}\ \bibnamefont
  {Lo}}, \bibinfo {author} {\bibfnamefont {H.-N.}\ \bibnamefont {Xiong}}, \
  and\ \bibinfo {author} {\bibfnamefont {W.-M.}\ \bibnamefont {Zhang}},\ }\href
  {https://www.nature.com/articles/srep09423} {\bibfield  {journal} {\bibinfo
  {journal} {Sci. Rep.}\ }\textbf {\bibinfo {volume} {5}},\ \bibinfo {pages}
  {9423} (\bibinfo {year} {2015})}\BibitemShut {NoStop}%
\bibitem [{\citenamefont {Li}\ \emph {et~al.}(2018)\citenamefont {Li},
  \citenamefont {Hall},\ and\ \citenamefont {Wiseman}}]{LHW18}%
  \BibitemOpen
  \bibfield  {author} {\bibinfo {author} {\bibfnamefont {L.}~\bibnamefont
  {Li}}, \bibinfo {author} {\bibfnamefont {M.~J.}\ \bibnamefont {Hall}}, \ and\
  \bibinfo {author} {\bibfnamefont {H.~M.}\ \bibnamefont {Wiseman}},\ }\href
  {http://www.sciencedirect.com/science/article/pii/S0370157318301601}
  {\bibfield  {journal} {\bibinfo  {journal} {Phys. Rep.}\ }\textbf {\bibinfo
  {volume} {759}},\ \bibinfo {pages} {1} (\bibinfo {year} {2018})}\BibitemShut
  {NoStop}%
\bibitem [{\citenamefont {Breuer}\ \emph {et~al.}(2016)\citenamefont {Breuer},
  \citenamefont {Laine}, \citenamefont {Piilo},\ and\ \citenamefont
  {Vacchini}}]{BLP16}%
  \BibitemOpen
  \bibfield  {author} {\bibinfo {author} {\bibfnamefont {H.-P.}\ \bibnamefont
  {Breuer}}, \bibinfo {author} {\bibfnamefont {E.-M.}\ \bibnamefont {Laine}},
  \bibinfo {author} {\bibfnamefont {J.}~\bibnamefont {Piilo}}, \ and\ \bibinfo
  {author} {\bibfnamefont {B.}~\bibnamefont {Vacchini}},\ }\href
  {https://journals.aps.org/rmp/abstract/10.1103/RevModPhys.88.021002}
  {\bibfield  {journal} {\bibinfo  {journal} {Rev. Mod. Phys.}\ }\textbf
  {\bibinfo {volume} {88}},\ \bibinfo {pages} {021002} (\bibinfo {year}
  {2016})}\BibitemShut {NoStop}%
\bibitem [{\citenamefont {Rivas}\ \emph {et~al.}(2014)\citenamefont {Rivas},
  \citenamefont {Huelga},\ and\ \citenamefont {Plenio}}]{RHP14}%
  \BibitemOpen
  \bibfield  {author} {\bibinfo {author} {\bibfnamefont {{\'A}.}~\bibnamefont
  {Rivas}}, \bibinfo {author} {\bibfnamefont {S.~F.}\ \bibnamefont {Huelga}}, \
  and\ \bibinfo {author} {\bibfnamefont {M.~B.}\ \bibnamefont {Plenio}},\
  }\href {Quantum non-Markovianity: characterization, quantification and
  detection} {\bibfield  {journal} {\bibinfo  {journal} {Rep. Prog. Phys.}\
  }\textbf {\bibinfo {volume} {77}},\ \bibinfo {pages} {094001} (\bibinfo
  {year} {2014})}\BibitemShut {NoStop}%
\bibitem [{\citenamefont {Heyl}(2018)}]{H18}%
  \BibitemOpen
  \bibfield  {author} {\bibinfo {author} {\bibfnamefont {M.}~\bibnamefont
  {Heyl}},\ }\href
  {https://iopscience.iop.org/article/10.1088/1361-6633/aaaf9a/meta} {\bibfield
   {journal} {\bibinfo  {journal} {Rep. Prog. Phys.}\ }\textbf {\bibinfo
  {volume} {81}},\ \bibinfo {pages} {054001} (\bibinfo {year}
  {2018})}\BibitemShut {NoStop}%
\bibitem [{\citenamefont {Trotzky}\ \emph {et~al.}(2012)\citenamefont
  {Trotzky}, \citenamefont {Chen}, \citenamefont {Flesch}, \citenamefont
  {McCulloch}, \citenamefont {Schollw{\"o}ck}, \citenamefont {Eisert},\ and\
  \citenamefont {Bloch}}]{TCF12}%
  \BibitemOpen
  \bibfield  {author} {\bibinfo {author} {\bibfnamefont {S.}~\bibnamefont
  {Trotzky}}, \bibinfo {author} {\bibfnamefont {Y.-A.}\ \bibnamefont {Chen}},
  \bibinfo {author} {\bibfnamefont {A.}~\bibnamefont {Flesch}}, \bibinfo
  {author} {\bibfnamefont {I.~P.}\ \bibnamefont {McCulloch}}, \bibinfo {author}
  {\bibfnamefont {U.}~\bibnamefont {Schollw{\"o}ck}}, \bibinfo {author}
  {\bibfnamefont {J.}~\bibnamefont {Eisert}}, \ and\ \bibinfo {author}
  {\bibfnamefont {I.}~\bibnamefont {Bloch}},\ }\href
  {https://www.nature.com/articles/nphys2232} {\bibfield  {journal} {\bibinfo
  {journal} {Nat. Phys.}\ }\textbf {\bibinfo {volume} {8}},\ \bibinfo {pages}
  {325} (\bibinfo {year} {2012})}\BibitemShut {NoStop}%
\bibitem [{\citenamefont {Gring}\ \emph {et~al.}(2012)\citenamefont {Gring},
  \citenamefont {Kuhnert}, \citenamefont {Langen}, \citenamefont {Kitagawa},
  \citenamefont {Rauer}, \citenamefont {Schreitl}, \citenamefont {Mazets},
  \citenamefont {Smith}, \citenamefont {Demler},\ and\ \citenamefont
  {Schmiedmayer}}]{GKL12}%
  \BibitemOpen
  \bibfield  {author} {\bibinfo {author} {\bibfnamefont {M.}~\bibnamefont
  {Gring}}, \bibinfo {author} {\bibfnamefont {M.}~\bibnamefont {Kuhnert}},
  \bibinfo {author} {\bibfnamefont {T.}~\bibnamefont {Langen}}, \bibinfo
  {author} {\bibfnamefont {T.}~\bibnamefont {Kitagawa}}, \bibinfo {author}
  {\bibfnamefont {B.}~\bibnamefont {Rauer}}, \bibinfo {author} {\bibfnamefont
  {M.}~\bibnamefont {Schreitl}}, \bibinfo {author} {\bibfnamefont
  {I.}~\bibnamefont {Mazets}}, \bibinfo {author} {\bibfnamefont {D.~A.}\
  \bibnamefont {Smith}}, \bibinfo {author} {\bibfnamefont {E.}~\bibnamefont
  {Demler}}, \ and\ \bibinfo {author} {\bibfnamefont {J.}~\bibnamefont
  {Schmiedmayer}},\ }\href
  {https://science.sciencemag.org/content/337/6100/1318} {\bibfield  {journal}
  {\bibinfo  {journal} {Science}\ }\textbf {\bibinfo {volume} {337}},\ \bibinfo
  {pages} {1318} (\bibinfo {year} {2012})}\BibitemShut {NoStop}%
\bibitem [{\citenamefont {Deutsch}(1991)}]{D91}%
  \BibitemOpen
  \bibfield  {author} {\bibinfo {author} {\bibfnamefont {J.~M.}\ \bibnamefont
  {Deutsch}},\ }\href {https://link.aps.org/doi/10.1103/PhysRevA.43.2046}
  {\bibfield  {journal} {\bibinfo  {journal} {Phys. Rev. A}\ }\textbf {\bibinfo
  {volume} {43}},\ \bibinfo {pages} {2046} (\bibinfo {year}
  {1991})}\BibitemShut {NoStop}%
\bibitem [{\citenamefont {Kosloff}(2013)}]{K13}%
  \BibitemOpen
  \bibfield  {author} {\bibinfo {author} {\bibfnamefont {R.}~\bibnamefont
  {Kosloff}},\ }\href {https://www.mdpi.com/1099-4300/15/6/2100/htm} {\bibfield
   {journal} {\bibinfo  {journal} {Entropy}\ }\textbf {\bibinfo {volume}
  {15}},\ \bibinfo {pages} {2100} (\bibinfo {year} {2013})}\BibitemShut
  {NoStop}%
\bibitem [{\citenamefont {Ali}\ \emph {et~al.}(2018)\citenamefont {Ali},
  \citenamefont {Huang}, \citenamefont {Zhang} \emph {et~al.}}]{AHZ18}%
  \BibitemOpen
  \bibfield  {author} {\bibinfo {author} {\bibfnamefont {M.}~\bibnamefont
  {Ali}}, \bibinfo {author} {\bibfnamefont {W.-M.}\ \bibnamefont {Huang}},
  \bibinfo {author} {\bibfnamefont {W.-M.}\ \bibnamefont {Zhang}},  \emph
  {et~al.},\ }\href {https://arxiv.org/abs/1803.04658} {\bibfield  {journal}
  {\bibinfo  {journal} {arXiv preprint arXiv:1803.04658}\ } (\bibinfo {year}
  {2018})}\BibitemShut {NoStop}%
\bibitem [{\citenamefont {Xiong}\ \emph {et~al.}(2015)\citenamefont {Xiong},
  \citenamefont {Lo}, \citenamefont {Zhang}, \citenamefont {Nori} \emph
  {et~al.}}]{XLZ15}%
  \BibitemOpen
  \bibfield  {author} {\bibinfo {author} {\bibfnamefont {H.-N.}\ \bibnamefont
  {Xiong}}, \bibinfo {author} {\bibfnamefont {P.-Y.}\ \bibnamefont {Lo}},
  \bibinfo {author} {\bibfnamefont {W.-M.}\ \bibnamefont {Zhang}}, \bibinfo
  {author} {\bibfnamefont {F.}~\bibnamefont {Nori}},  \emph {et~al.},\ }\href
  {https://www.nature.com/articles/srep13353} {\bibfield  {journal} {\bibinfo
  {journal} {Sci. Rep.}\ }\textbf {\bibinfo {volume} {5}},\ \bibinfo {pages}
  {13353} (\bibinfo {year} {2015})}\BibitemShut {NoStop}%
\bibitem [{\citenamefont {Srednicki}(1994)}]{S94}%
  \BibitemOpen
  \bibfield  {author} {\bibinfo {author} {\bibfnamefont {M.}~\bibnamefont
  {Srednicki}},\ }\href {https://link.aps.org/doi/10.1103/PhysRevE.50.888}
  {\bibfield  {journal} {\bibinfo  {journal} {Phys. Rev. E}\ }\textbf {\bibinfo
  {volume} {50}},\ \bibinfo {pages} {888} (\bibinfo {year} {1994})}\BibitemShut
  {NoStop}%
\bibitem [{\citenamefont {Linden}\ \emph {et~al.}(2009)\citenamefont {Linden},
  \citenamefont {Popescu}, \citenamefont {Short},\ and\ \citenamefont
  {Winter}}]{LPS09}%
  \BibitemOpen
  \bibfield  {author} {\bibinfo {author} {\bibfnamefont {N.}~\bibnamefont
  {Linden}}, \bibinfo {author} {\bibfnamefont {S.}~\bibnamefont {Popescu}},
  \bibinfo {author} {\bibfnamefont {A.~J.}\ \bibnamefont {Short}}, \ and\
  \bibinfo {author} {\bibfnamefont {A.}~\bibnamefont {Winter}},\ }\href
  {https://link.aps.org/doi/10.1103/PhysRevE.79.061103} {\bibfield  {journal}
  {\bibinfo  {journal} {Phys. Rev. E}\ }\textbf {\bibinfo {volume} {79}},\
  \bibinfo {pages} {061103} (\bibinfo {year} {2009})}\BibitemShut {NoStop}%
\bibitem [{\citenamefont {Short}\ and\ \citenamefont {Farrelly}(2012)}]{SF12}%
  \BibitemOpen
  \bibfield  {author} {\bibinfo {author} {\bibfnamefont {A.~J.}\ \bibnamefont
  {Short}}\ and\ \bibinfo {author} {\bibfnamefont {T.~C.}\ \bibnamefont
  {Farrelly}},\ }\href
  {https://iopscience.iop.org/article/10.1088/1367-2630/14/1/013063/meta}
  {\bibfield  {journal} {\bibinfo  {journal} {New J. Phys.}\ }\textbf {\bibinfo
  {volume} {14}},\ \bibinfo {pages} {013063} (\bibinfo {year}
  {2012})}\BibitemShut {NoStop}%
\bibitem [{\citenamefont {Reimann}(2008)}]{R08}%
  \BibitemOpen
  \bibfield  {author} {\bibinfo {author} {\bibfnamefont {P.}~\bibnamefont
  {Reimann}},\ }\href {https://link.aps.org/doi/10.1103/PhysRevLett.101.190403}
  {\bibfield  {journal} {\bibinfo  {journal} {Phys. Rev. Lett.}\ }\textbf
  {\bibinfo {volume} {101}},\ \bibinfo {pages} {190403} (\bibinfo {year}
  {2008})}\BibitemShut {NoStop}%
\bibitem [{\citenamefont {Rigol}(2009)}]{R09}%
  \BibitemOpen
  \bibfield  {author} {\bibinfo {author} {\bibfnamefont {M.}~\bibnamefont
  {Rigol}},\ }\href {https://link.aps.org/doi/10.1103/PhysRevLett.103.100403}
  {\bibfield  {journal} {\bibinfo  {journal} {Phys. Rev. Lett.}\ }\textbf
  {\bibinfo {volume} {103}},\ \bibinfo {pages} {100403} (\bibinfo {year}
  {2009})}\BibitemShut {NoStop}%
\bibitem [{\citenamefont {Polkovnikov}\ \emph {et~al.}(2011)\citenamefont
  {Polkovnikov}, \citenamefont {Sengupta}, \citenamefont {Silva},\ and\
  \citenamefont {Vengalattore}}]{PSSV11}%
  \BibitemOpen
  \bibfield  {author} {\bibinfo {author} {\bibfnamefont {A.}~\bibnamefont
  {Polkovnikov}}, \bibinfo {author} {\bibfnamefont {K.}~\bibnamefont
  {Sengupta}}, \bibinfo {author} {\bibfnamefont {A.}~\bibnamefont {Silva}}, \
  and\ \bibinfo {author} {\bibfnamefont {M.}~\bibnamefont {Vengalattore}},\
  }\href {https://link.aps.org/doi/10.1103/RevModPhys.83.863} {\bibfield
  {journal} {\bibinfo  {journal} {Rev. Mod. Phys.}\ }\textbf {\bibinfo {volume}
  {83}},\ \bibinfo {pages} {863} (\bibinfo {year} {2011})}\BibitemShut
  {NoStop}%
\bibitem [{\citenamefont {Cazalilla}\ and\ \citenamefont {Rigol}(2010)}]{CR10}%
  \BibitemOpen
  \bibfield  {author} {\bibinfo {author} {\bibfnamefont {M.}~\bibnamefont
  {Cazalilla}}\ and\ \bibinfo {author} {\bibfnamefont {M.}~\bibnamefont
  {Rigol}},\ }\href
  {https://iopscience.iop.org/article/10.1088/1367-2630/12/5/055006/meta}
  {\bibfield  {journal} {\bibinfo  {journal} {New J. Phys.}\ }\textbf {\bibinfo
  {volume} {12}},\ \bibinfo {pages} {055006} (\bibinfo {year}
  {2010})}\BibitemShut {NoStop}%
\bibitem [{\citenamefont {Hsiang}\ \emph {et~al.}(2018)\citenamefont {Hsiang},
  \citenamefont {Chou}, \citenamefont {Suba{\c{s}}{\i}},\ and\ \citenamefont
  {Hu}}]{HCSH18}%
  \BibitemOpen
  \bibfield  {author} {\bibinfo {author} {\bibfnamefont {J.-T.}\ \bibnamefont
  {Hsiang}}, \bibinfo {author} {\bibfnamefont {C.~H.}\ \bibnamefont {Chou}},
  \bibinfo {author} {\bibfnamefont {Y.}~\bibnamefont {Suba{\c{s}}{\i}}}, \ and\
  \bibinfo {author} {\bibfnamefont {B.~L.}\ \bibnamefont {Hu}},\ }\href
  {https://journals.aps.org/pre/abstract/10.1103/PhysRevE.97.012135} {\bibfield
   {journal} {\bibinfo  {journal} {Phys. Rev. E}\ }\textbf {\bibinfo {volume}
  {97}},\ \bibinfo {pages} {012135} (\bibinfo {year} {2018})}\BibitemShut
  {NoStop}%
\bibitem [{\citenamefont {Callen}(1998)}]{C98}%
  \BibitemOpen
  \bibfield  {author} {\bibinfo {author} {\bibfnamefont {H.~B.}\ \bibnamefont
  {Callen}},\ }\href@noop {} {\enquote {\bibinfo {title} {Thermodynamics and an
  introduction to thermostatistics},}\ } (\bibinfo {year} {1998})\BibitemShut
  {NoStop}%
\bibitem [{\citenamefont {Leggett}\ \emph {et~al.}(1987)\citenamefont
  {Leggett}, \citenamefont {Chakravarty}, \citenamefont {Dorsey}, \citenamefont
  {Fisher}, \citenamefont {Garg},\ and\ \citenamefont {Zwerger}}]{LCD87}%
  \BibitemOpen
  \bibfield  {author} {\bibinfo {author} {\bibfnamefont {A.~J.}\ \bibnamefont
  {Leggett}}, \bibinfo {author} {\bibfnamefont {S.}~\bibnamefont
  {Chakravarty}}, \bibinfo {author} {\bibfnamefont {A.~T.}\ \bibnamefont
  {Dorsey}}, \bibinfo {author} {\bibfnamefont {M.~P.~A.}\ \bibnamefont
  {Fisher}}, \bibinfo {author} {\bibfnamefont {A.}~\bibnamefont {Garg}}, \ and\
  \bibinfo {author} {\bibfnamefont {W.}~\bibnamefont {Zwerger}},\ }\href
  {https://link.aps.org/doi/10.1103/RevModPhys.59.1} {\bibfield  {journal}
  {\bibinfo  {journal} {Rev. Mod. Phys.}\ }\textbf {\bibinfo {volume} {59}},\
  \bibinfo {pages} {1} (\bibinfo {year} {1987})}\BibitemShut {NoStop}%
\bibitem [{\citenamefont {Zhang}\ \emph {et~al.}(1990)\citenamefont {Zhang},
  \citenamefont {Feng},\ and\ \citenamefont {Gilmore}}]{ZFG90}%
  \BibitemOpen
  \bibfield  {author} {\bibinfo {author} {\bibfnamefont {W.-M.}\ \bibnamefont
  {Zhang}}, \bibinfo {author} {\bibfnamefont {D.~H.}\ \bibnamefont {Feng}}, \
  and\ \bibinfo {author} {\bibfnamefont {R.}~\bibnamefont {Gilmore}},\ }\href
  {https://link.aps.org/doi/10.1103/RevModPhys.62.867} {\bibfield  {journal}
  {\bibinfo  {journal} {Rev. Mod. Phys.}\ }\textbf {\bibinfo {volume} {62}},\
  \bibinfo {pages} {867} (\bibinfo {year} {1990})}\BibitemShut {NoStop}%
\bibitem [{\citenamefont {Kadanoff}(2018)}]{K18}%
  \BibitemOpen
  \bibfield  {author} {\bibinfo {author} {\bibfnamefont {L.~P.}\ \bibnamefont
  {Kadanoff}},\ }\href@noop {} {\emph {\bibinfo {title} {Quantum Statistical
  Mechanics}}}\ (\bibinfo  {publisher} {CRC Press},\ \bibinfo {year}
  {2018})\BibitemShut {NoStop}%
\bibitem [{\citenamefont {Tillmann}\ \emph {et~al.}(2013)\citenamefont
  {Tillmann}, \citenamefont {Daki{\'c}}, \citenamefont {Heilmann},
  \citenamefont {Nolte}, \citenamefont {Szameit},\ and\ \citenamefont
  {Walther}}]{TDH13}%
  \BibitemOpen
  \bibfield  {author} {\bibinfo {author} {\bibfnamefont {M.}~\bibnamefont
  {Tillmann}}, \bibinfo {author} {\bibfnamefont {B.}~\bibnamefont {Daki{\'c}}},
  \bibinfo {author} {\bibfnamefont {R.}~\bibnamefont {Heilmann}}, \bibinfo
  {author} {\bibfnamefont {S.}~\bibnamefont {Nolte}}, \bibinfo {author}
  {\bibfnamefont {A.}~\bibnamefont {Szameit}}, \ and\ \bibinfo {author}
  {\bibfnamefont {P.}~\bibnamefont {Walther}},\ }\href
  {https://www.nature.com/articles/nphoton.2013.102} {\bibfield  {journal}
  {\bibinfo  {journal} {Nat. Photon.}\ }\textbf {\bibinfo {volume} {7}},\
  \bibinfo {pages} {540} (\bibinfo {year} {2013})}\BibitemShut {NoStop}%
\bibitem [{\citenamefont {Aaronson}\ and\ \citenamefont
  {Arkhipov}(2011)}]{AA11}%
  \BibitemOpen
  \bibfield  {author} {\bibinfo {author} {\bibfnamefont {S.}~\bibnamefont
  {Aaronson}}\ and\ \bibinfo {author} {\bibfnamefont {A.}~\bibnamefont
  {Arkhipov}},\ }in\ \href@noop {} {\emph {\bibinfo {booktitle} {Proceedings of
  the Forty-Third Annual ACM Symposium on Theory of Computing}}}\ (\bibinfo
  {organization} {ACM},\ \bibinfo {year} {2011})\ pp.\ \bibinfo {pages}
  {333--342}\BibitemShut {NoStop}%
\bibitem [{\citenamefont {Feynman}\ \emph {et~al.}(2011)\citenamefont
  {Feynman}, \citenamefont {Leighton},\ and\ \citenamefont {Sands}}]{F11}%
  \BibitemOpen
  \bibfield  {author} {\bibinfo {author} {\bibfnamefont {R.~P.}\ \bibnamefont
  {Feynman}}, \bibinfo {author} {\bibfnamefont {R.~B.}\ \bibnamefont
  {Leighton}}, \ and\ \bibinfo {author} {\bibfnamefont {M.}~\bibnamefont
  {Sands}},\ }\href@noop {} {\emph {\bibinfo {title} {The Feynman Lectures on
  Physics}}},\ Vol.~\bibinfo {volume} {3}\ (\bibinfo  {publisher} {Basic
  books},\ \bibinfo {year} {2011})\BibitemShut {NoStop}%
\bibitem [{\citenamefont {Sharma}\ and\ \citenamefont {Rabani}(2015)}]{SR15}%
  \BibitemOpen
  \bibfield  {author} {\bibinfo {author} {\bibfnamefont {A.}~\bibnamefont
  {Sharma}}\ and\ \bibinfo {author} {\bibfnamefont {E.}~\bibnamefont
  {Rabani}},\ }\href
  {https://journals.aps.org/prb/abstract/10.1103/PhysRevB.91.085121} {\bibfield
   {journal} {\bibinfo  {journal} {Phys. Rev. B}\ }\textbf {\bibinfo {volume}
  {91}},\ \bibinfo {pages} {085121} (\bibinfo {year} {2015})}\BibitemShut
  {NoStop}%
\bibitem [{\citenamefont {Kamenev}\ and\ \citenamefont
  {Levchenko}(2009)}]{KL09}%
  \BibitemOpen
  \bibfield  {author} {\bibinfo {author} {\bibfnamefont {A.}~\bibnamefont
  {Kamenev}}\ and\ \bibinfo {author} {\bibfnamefont {A.}~\bibnamefont
  {Levchenko}},\ }\href
  {https://www.tandfonline.com/doi/full/10.1080/00018730902850504} {\bibfield
  {journal} {\bibinfo  {journal} {Adv. Phys.}\ } (\bibinfo {year}
  {2009})}\BibitemShut {NoStop}%
\end{thebibliography}%

\end{document}